\documentclass[12pt]{article}
\usepackage{graphicx}
\usepackage{amssymb,amsmath,latexsym}
\usepackage{amsthm}
\newtheorem{defi}{Definition}
\newtheorem{remark}{Remark}
\newtheorem{theom}{Theorem}

\newtheorem{prop}{Proposition}
\newtheorem{proper}{Property}
\def\Bot{\!\!\! _{{\displaystyle\bot}}}
\def\bmth#1{{\mbox{\boldmath$#1$}}}
\def\slacs#1{\setlength{\arraycolsep}{#1}}
\def\sltcs#1{\setlength{\tabcolsep}{#1}}

\def\scsty{\scriptstyle}
\def\be{\begin{equation}}
\def\ee{\end{equation}}
\def\bea{\begin{eqnarray}}
\def\eea{\end{eqnarray}}
\def\bei{\begin{itemize}}
\def\eni{\end{itemize}}
\def\ben{\begin{enumerate}}
\def\ene{\end{enumerate}}
\def\disty{\displaystyle}

\def\D{\,{\rm d}}
\def\L{{\rm L}}
\def\S{{\rm S}}

\def\ba{\begin{array}}
\def\ea{\end{array}}
\def\R{\mathbb{R}}
\def\Q{\mathbb{Q}}
\def\Z{\mathbb{Z}}
\def\N{\mathbb{N}}

\def\LR{L^{2}\mathbb{(R)}}
\def\ZB{\Z_{\beta}}
\def\BP{\mathcal B}
\def\BPM{{\mathcal B}_-}
\def\BPP{{\mathcal B}_+}
\def\FT{{\mathcal F}_{\tau}}
\def\FB{{\mathcal F}_{\beta}}

\def\T{\mathcal T}
\def\x{\mathbf{x}}
\def\LA{\Lambda}

\author{Miroslav Andrle$^{\dag\ddag}$, \v{C}estm\'{\i}r
Burd\'{\i}k$^{\ddag}$, and Jean-Pierre Gazeau$^{\P}$}
\date{
{\it\small $^{\dag}$NCRG, Aston University, Aston Triangle, Birmingham,}\\
{\it \small  B4 7ET, United
 Kingdom}\\
{\it\small $^{\ddag}$Dep. of Mathematics,}
{\it\small FNSPE-CTU, Trojanova 13,}\\
{\it\small 120\,00 Prague 2, Czech Republic}\\{\small and}\\
{\it\small $^{\P}$LPTMC,}
{\it\small Box 7020, Universit\'e Paris 7--Denis Diderot,}\\
{\it\small 75251 Paris Cedex 05, France}\\
{\it\small  E-mail: \tt gazeau@ccr.jussieu.fr}
}

\title{\sc Bernuau spline wavelets and Sturmian sequences}

\begin{document}
\maketitle
\begin{abstract}

  We present spline wavelets of class $C^n(\R)$
supported by sequences of aperiodic
discretizations of $\R$.  The construction is based on multiresolution analysis
recently elaborated by G. Bernuau. At a given scale, we consider
discretizations  that are  sets of left-hand ends of  tiles in a
self-similar tiling of the real line with finite local complexity. Corresponding
 tilings are determined by two-letter Sturmian substitution
sequences. 
We illustrate the construction with  examples having quadratic
Pisot-Vijayaraghavan units (like
$\tau = (1 + \sqrt{5})/2$ or $\tau^2 = (3 + \sqrt{5})/2$) as scaling
factor. In particular, we present a comprehensive analysis of the Fibonacci chain and give the analytic form of related scaling functions and wavelets.  We also give some hints for the construction of  multidimensional spline wavelets based on stone-inflation tilings in arbitrary dimension.

\end{abstract}

\section{Introduction}
Under the name ``wavelet'' is commonly understood a function
$\psi(x)\in\LR$ such that the family of functions
$\psi_{j,k}(x)=2^{j/2}\psi(2^j x-k)$ for $j,k\in\Z$ forms an
orthonormal (or at least a Riesz) basis for $\LR$. A function generating through dilatations  and
translations an orthonormal basis for $\LR$  can be found through
a ``multiresolution analysis of $\LR$'' (shortly MRA), a method
settled by S.~Mallat \cite{mallat}. The dilatation factor is usually
$\theta=2$.   Indeed,  the construction of a wavelet basis within the
MRA framework relies on the fact that the lattices $2^{-j}\Z$
are increasing for the inclusion. This property is preserved only
when $\theta$ is an integer. Then one can raise a natural question:  what about choosing another number $\theta$
as a scaling factor? A first answer is given
in the work of P.~Auscher \cite{auscher}. The following problem was considered:
given a real number $\theta>1$, does there exist a finite set
$\{\psi_1,\psi_2,\dots,\psi_\ell\}$ of functions in $\LR$ such
that the family $\theta^{j/2}\psi_i(\theta^j x-k),\ j,k\in\Z,\ 1 \leq i\leq
\ell,$ is an orthonormal  basis for $\LR$?  The author then proved
 that a basis of this type exists if  $\theta$ is
a rational number. More precisely, for $\theta=p/q>1$, $p$ and $q$
being relatively prime integers, there exists a set of $p-q$
wavelet functions satisfying the previous condition. It was still an open  question whether there exists another
generalization of wavelet basis with
 an irrational scaling factor.
 On the other hand, in 1992 Buhmann and Micchelli  \cite{bumi} proposed  a
construction of a wavelet spline basis corresponding to
non-uniform and non-self-similar knot sequences. Further \cite{dala1,dala2,waya} studies have been recently devoted to this problem in higher dimension, like answering the question of characterizing functions $\psi$,
dilation sets $\mathcal{D}$, and translation sets $\mathcal{T}$, such that 
$\left\{ \vert \mathrm{det}(D) \vert^{\frac{1}{2}} \psi(Dx - \lambda) \, | \, D \in \mathcal{D}, \lambda \in \mathcal{T}  \right\}$ forms an orthonormal basis for 
$\LR$ \cite{waya}. As a matter of fact, it was proved in \cite{dala2} that for any real
expansive (\textit{i.e.}  all eigenvalues have modulus greater than one) $d\times d$ matrix $A$, there exists a measurable  $E \subset  \R^d$ such that $\left\{\vert \det{A}\vert^{j/2} {\bar{\hat{\chi}}}_E (A^j x - l), \ j \in \Z, l\in \Z^d  \right\}$ forms an orthonormal basis for $L^2(\R^d)$.

In 1996, a Haar wavelet basis with
an algebraic irrational scaling factor and only one generating wavelet
$\psi(x)$ was given in
\cite{gazpat}. This wavelet basis  of $\LR$ lives on the
nested sets $\tau ^j \Z_\tau$ having the following structure:
$\tau^{j/2}\psi(\tau^j x-\lambda),\ j\in\Z,\ \lambda\in\Lambda$,
where $\tau=(1+\sqrt{5})/2$ is the golden mean. Set  $\Z_\tau$
means set of ``tau-integers'' and  set $\Lambda$ is the set of admissible translations. One  outcome of the latter work is that admissible  translations are generically irrational for an irrational scaling factor. This seems to be a common feature to aperiodic translational sets. 
In 1998 G. Bernuau \cite{bern1, bern2} settled  a  general construction of spline
wavelets living on locally finite (more precisely with finite local complexity) and self-similar Delaunay (or Delone) sets. His approach was mainly inspired by important results previously obtained by P.G. Lemari\'e-Rieusset \cite{lem} in the more general framework of stratified nilpotent Lie groups.
We learn from these last works  that  irrational
factors combined with specific properties of sequence of discretizations imply finitely many generating wavelets and that we also
need more functions named ``scaling functions'' in an appropriate
modification of MRA. Actually the number of spline scaling
functions and spline wavelets does not depend only on the scaling
factor but also on the polynomial order of these functions. 

In a previous letter \cite{abug}, we have presented a definition of
multiresolution analysis for an infinite sequence $\cdots
\subset\mathcal{F}/\tau^{2j-2}\subset\mathcal{F}/\tau^{2j}\subset
\mathcal{F}/\tau^{2j+2}\subset\cdots $ of aperiodic discretizations
of $\mathbb{R}$. Corresponding wavelets have been defined and the
elementary Haar example was given as an illustration of the
method. Starting from the Haar case, one could be attempted to explore less trivial
examples in which \underline{compact} support and \underline{
orthogonality} are required for wavelet families of class $C^{n}$, $n\geq 0$.
This is a strong constraint for such {\it Daubechies-like} wavelets which live on
aperiodic discretizations, and  their existence is  not
guaranteed at the moment. Our present strategy is to drop out the
orthogonality condition and to rather explore certain well-known quasiperiodic
counterparts of the dyadic spline wavelets.  In the present paper, we shall closely follow
the procedure rigorously settled  by G.~Bernuau. Let us  briefly summarize the general setting of the
Bernuau construction.
Let $\Lambda \subset \R$ be a Delaunay point set in the real line. By Delaunay we mean that $\Lambda$
is {\it uniformly discrete} (the distances between any pair of points in $\Lambda$
are greater than a fixed
$r >0$)
\underline{and} {\it relatively dense} (there exists $R>0$ such that $\R$ is covered by intervals of
length $2R$ centered at points of $\Lambda$). In addition to this Delaunay structure, we demand that
the set $\Lambda$ be
\ben
\item \emph{self-similar}: there exists a number $\theta > 1 $  ({\it inflation factor}) such that
\begin{equation}
\theta \Lambda \subset \Lambda,
\end{equation}
\item \emph{with finite local complexity}: for all $R > 0$, the point set
\begin{equation}
\bigcup_{\lambda \in \Lambda}\left\{(\Lambda - \lambda) \cap (-R,R) \right\}
\end{equation}
is finite. This means that local environments of points in
$\Lambda$ are not  different in infinite fashions. \ene Typically,
such sets  $\Lambda$ are mathematical models for one-dimensional
structures having a long-range order, like quasicrystals. Our aim
here is to construct a Riesz basis of $L^2 (\R)$, the elements of
which are of the affine wavelet type:
\begin{equation}
\theta^{j/2} \psi_{\kappa} (\theta^j x - \kappa), \ j\in\Z, \
\kappa \in \theta^{-1}\Lambda, \ \mbox{succ}(\kappa) \notin \Lambda,
\end{equation}
where $\mbox{succ}(\kappa) $ is the nearest  right neighbour of $\kappa$ in the set $\theta^{-1} \Lambda$,
and for which the set
$
\left\{ \psi_{\kappa}(x)\right\}
$
of {\it mother wavelets}
is finite and the simplest possible.

At this point, we recall that
$(v_n)$ is a Riesz basis of a separable Hilbert space $V$ if and
only if each $v \in V$ can be expressed uniquely as $ v = \sum_n
a_n v_n $ and there exist positive constants $A$ and $B,0<A\leq B$, such
that
\begin{equation}
A \sum_n \vert a_n \vert^2  \leq \Big\Vert\sum_n a_n v_n
\Big\Vert^2 \leq B \sum_n \vert a_n \vert^2
\end{equation}
for all sequence of scalars $a_n$. We can say that the $v_n$'s are strongly linearly independent.

In the next section, we shall present a survey of preliminary
results concerning  the space of splines based on $\Lambda$ and at
the heart of the construction of wavelets and  corresponding
multiresolution analysis. In Section 3, we recall the Bernuau  theorems about the
existence and the characterization of the wavelet basis itself. Section 4 is devoted to the description of
Delaunay sets based on two-letter Sturmian substitution sequences.
These sets have as scaling factor special  algebraic integers,
with generic symbol $\beta$, and named  {\it quadratic
Pisot-Vijayaraghavan units} or more simply {UPV$ _2$}. Among the
latter one finds those numbers which are of interest in
quasicrystalline studies: $\tau = (1 + \sqrt{5})/2 $, $\tau^2 = (3
+ \sqrt{5})/2$ (for pentagonal and decagonal cases), $\omega = 1 +
\sqrt2 $, $\omega^2 = 3 + 2\sqrt2$ (for octogonal case), and
$\delta = 2 + \sqrt3 $ (for dodecagonal case). In this context, we
shall consider two types of point sets: the $\beta$-integers $\ZB$
and certain {\it model sets} included in the latter. Original results are given in Sections 5--7.
 Haar wavelets living on  $\beta$-integer sets with corresponding scaling equations are given in Section 5. Section 6 is devoted to the study of one important example of model set, namely the Fibonacci  chain, from a lexicographical point of view (see Proposition 6 and Properties 7-9) and prepares the section 7  in which
related splines and wavelets together with their scaling equations are explicitly constructed. In Section 8, we consider a class of self-similar tilings of $\R^d$ having the so-called stone-inflation property. For instance, we can find stone-inflation tilings among Penrose or triangle tilings of the plane. We just sketch the method of construction of Bernuau spline wavelets adapted to such tilings. 

\section{Space of splines for multiresolution analysis}

\subsection{The definition of the space  $\bmth{V_0^{(s)}(\Lambda)}$}
Any Delaunay set $\Lambda$ determines a space of splines of order $s$, $s\geq 2$, in the following way
\begin{defi}\label{v0}
Let $s\geq 2$.  Then ${V_0^{(s)}(\Lambda)}$ is the closed subspace of $\LR$ defined by
 ${V_0^{(s)}(\Lambda)}=\left\{f(x)\in \LR \,  \Big | \,\cfrac[l]{\D^s}{\D x^s}f(x)=\sum_{\lambda\in\Lambda}
a_{\lambda} \delta_{\lambda}\right\}.$
\end{defi}
An equivalent definition is given in terms of the restriction of
functions to intervals determined by consecutive elements of $\Lambda$. Suppose the latter is
defined by the increasing one-to-one map from $\Z$ : $\Z \ni n \rightarrow \lambda_n $, $\cdots
\lambda_{n-1} < \lambda_n < \lambda_{n +1} \cdots $. There results from Def.\,1 that
${V_0^{(s)}(\Lambda)}=\left\{f\in C^{s-2},
\ f\in\LR,
\  f_{\mid_{[\lambda_n,\lambda_{n+1}]}} \mbox{ is a polynomial of
degree }\leq s - 1 \right\}$.
Therefore, ${V_0^{(s)}(\Lambda)}$ is the space of splines of order $s$ with nodes in $\Lambda$. Let
us now give a classical result about the existence of a Riesz basis for ${V_0^{(s)}(\Lambda)}$
\cite{schum}.

\begin{theom} For all Delaunay  sets $\Lambda=\{\lambda_n\}_{n\in\Z}\subset\R$ and for all $s\geq 2$,
  there exists a Riesz basis $\{ B_{\lambda}^{(s)}, \lambda\in\Lambda\}$
  of ${V_0^{(s)}(\Lambda)}$.   The function  $B_{\lambda}^{(s)}$ (called B-spline)
   is the unique function in
   ${V_0^{(s)}(\Lambda)}$ satisfying the following conditions:
   \begin{enumerate}
   \item[(i)]{$\mathrm{supp}\ B_{\lambda}^{(s)}=[\lambda,\lambda']$, where $\lambda'\in\Lambda.$}
   \item[(ii)]{The interval $(\lambda,\lambda')$ contains exactly $s-1$ points of $\Lambda$.}
   \item[(iii)]{$\displaystyle\int_{\R}B_{\lambda}^{(s)}=\frac{\lambda'-\lambda}{s}.$}
   \end{enumerate}
\end{theom}
See \cite{schum} for proof.
Note that (i) and (ii) give precise information on the (compact) support of $B_{\lambda}^{(s)}$
whilst (iii) is a normalization condition.

\subsection{Construction of B-splines of order $\bmth{s,\ B^{(s)}_{\lambda}(x)}$}

The construction of $B^{(s)}_{\lambda}(x)$ can be carried out in various ways:
\begin{itemize}
\item{by  recurrence  \cite{ris}}, with the use of the formula valid for all $\lambda_n\in\Lambda$
\slacs{.6mm}
\begin{equation}
B^{(1)}_{\lambda_n}(x) = \left\{%
\begin{array}{ll}
   \ 1 & \quad \hbox{if $\lambda_n \leq x<\lambda_{n+1}$,} \\
   \ 0 & \quad\hbox{otherwise,} \\
\end{array}%
\right.
\end{equation}
\begin{equation}
B^{(s)}_{\lambda_n}(x)= \omega_{s,n} (x) B^{(s-1)}_{\lambda_n}(x) + \left(1 - \omega_{s,n + 1} (x)\right)
B^{(s-1)}_{\lambda_{n+1}}(x), \mbox{ for } s\geq 2
\end{equation}
$\mbox{with }\omega_{s,n} (x) =
 (x - \lambda_n)/(\lambda_{n + s - 1}-\lambda_n ),$
\item{by using the condition of minimal support, which means that $\forall f \in
V_0^{(s)}(\Lambda)$,
\begin{equation}
\mathrm{supp}\ f \subset \mathrm{supp}\ B_{\lambda}^{(s)} \Rightarrow f \propto  B_{\lambda}^{(s)},
\end{equation}}
\item{by inverse Fourier transform which gives the explicit form}
\begin{equation}
B_{\lambda_n}^{(s)} (x) \propto \mbox{{\cal F}ourier}^{-1}\left\lbrack \frac{1}{(i\xi)^s} \left(
\sum_{l=0}^{s} a_l (n) e^{-i\lambda_{n+l}\xi}
\right)\right\rbrack, \label{four}
\end{equation}
in which $\left\{ a_l (n) \right\} $ is the unique solution of the linear
system
\begin{equation} 
 \ba{l}\disty
\sum_{l=0}^{s} (\lambda_{n+l})^j x_l = 0, \ \ 0\leq j \leq s-1,\\
\disty\sum_{l=0}^{s} (\lambda_{n+l})^s x_l = \frac{(-1)^s}{s!}.
\ea
\end{equation}
\end{itemize}
Note that the above linear system is easily solved since it involves Vandermonde determinants:
\begin{equation}
a_l (n) = \frac{(-1)^l}{s!}\left\lbrack \prod_{0 \leq l' <l} (\lambda_{n+l} -
\lambda_{n+l'}) \prod_{ l <l' \leq s} (\lambda_{n+l'} -
\lambda_{n+l})\right\rbrack^{-1}.
\end{equation}
 From (\ref{four}) one can check that the $s^{\text{th}}$ derivative of $B_{\lambda_n}^{(s)} (x)$
is given as a finite linear superposition of Dirac masses located at the points
$\lambda_{n + l}, \, 0 \leq l \leq s$:
$$
\frac{\D^s}{\D x^s}B_{\lambda_n}^{(s)} (x) \propto \sum_{l=0}^s
a_l (n) \delta_{\lambda_{n + l}}.
$$
Since  the  Fourier transform of $B_{\lambda_n}^{(s)} (x
+ \lambda_n)$ depends on the $s$-tuple $(\lambda_{n + 1} -
\lambda_n, \dots, \lambda_{n + s} - \lambda_n ) $ only and 
since such $s$-tuples assume their values in a finite set for $n$ varying in
$\Z$ if $\Lambda$ has
 finite local complexity, we can assert the following:

\begin{prop}\label{prop1}
Let $\Lambda\subset\R$ be a Delaunay set of finite local complexity. Then the
 B-splines  of order $s$ based on $\Lambda$ are of the form
 $B^{(s)}_{\lambda}(x)=\phi_{\lambda}(x-\lambda),\ \lambda\in\Lambda$, where the set
  $\{\phi_{\lambda}(x),\ \lambda\in\Lambda\}$ is a finite set of
  functions with  compact support.
\end{prop}

Therefore, in the finite local complexity case, it is possible to partition the indexing set $\Z$ for $\Lambda$
into a finite set of equivalence classes $\bar{n}$, $\Z = \overset{\bar{q}}{\underset{\bar{n} =
\bar{0}}\cup}\bar{n}$ where class $\bar{n}$ is defined by
$$
\bar{n} = \left\{ k \in \Z \, \mid \, B_{\lambda_{k}}^{(s)} (x + \lambda_{k}) =
B_{\lambda_n}^{(s)} (x + \lambda_n) \ \forall x \right\}.
$$
Correspondingly, for a given $s$, the point set $\Lambda$ is partitioned into $\Lambda = \overset{\bar{q}}{\underset{\bar{n} =
\bar{0}}\cup} \Lambda_{\bar{n}}$ with $\Lambda_{\bar{n}} = \left\{
\lambda_k \in \LA \, | \, k\in \bar{n}
\right\} $. The relation 
$\lambda_k \in \Lambda_{\bar{n}}$ means that $\lambda_k$ and $\lambda_n$ are left-hand ends of identical $s$-letter words if we identify each interval $\left( \lambda_k , \lambda_{k+1}\right)$ with a letter of the 
allowed alphabet. To each class $\bar{n}$ is biunivocally associated the function $\phi_{\bar{n}} (x)
\equiv \phi_{\lambda_k} (x) = B_{\lambda_k}^{(s)} (x + \lambda_k), \ k \in \bar{n} $.  In this way,
the space $V_0^{(s)}(\Lambda)$ decomposes into the  direct sum
\begin{equation}
V_0^{(s)}(\Lambda) = \overset{\bar{q}}{\underset{\bar{n} =
\bar{0}}\bigoplus} V_{0, \bar{n}},
\end{equation}
where $V_{0, \bar{n}} $ is the closure of the linear span of the functions  $\phi_{\bar{n}} (x -
\lambda_k), \ k \in  \bar{n}$.

\subsection{Self-similarity and multiresolution analysis}

 Let $\Lambda$ be a Delaunay set of finite local complexity and self-similar
with inflation factor $\theta > 1$ ($\theta\Lambda\subset\Lambda$).
 Changing the scale allows us to define subspaces ${V_j^{(s)}(\Lambda)},\ j\in\Z.$
\begin{defi}\label{Vj}
${V_j^{(s)}(\Lambda)}=\left\{f(x)\in \LR\ \mid \  \frac{\D^s}{\D x^s}f(x)=\sum_{\lambda\in\Lambda}
a_{\lambda} \delta_{\theta^{-j}\lambda} \right\}.$
\end{defi}
We  now have at our disposal an inductive chain of spaces allowing analysis at any scale. More
precisely, with the above notations,
\begin{prop}
The sequence of  subspaces  $(V_j^{(s)}(\Lambda))_{j\in
\mathbb{Z}}$ is a $\theta$-multiresolution analysis of $L^{2}\mathbb{(R})$, i.e.
\begin{enumerate}\vspace{-2mm}\label{multiresolution}
\item[(i)]{for any $j\in\Z, \ V_j^{(s)}(\Lambda)$ is a closed subspace of $L^{2}\mathbb{(R})$,}\vspace{-2mm}
\item[(ii)]{$\cdots\subset V_{-1}^{(s)}(\Lambda)\subset V_0^{(s)}(\Lambda)\subset V_1^{(s)}(\Lambda)
\subset\cdots$,}\vspace{-2mm}
\item[(iii)]{$\bigcup_{j\in\Z}V_j^{(s)}(\Lambda)$ is dense in $L^{2}\mathbb{(R})$,}\vspace{-2mm}
\item[(iv)]{$\bigcap_{j\in\Z}V_j^{(s)}(\Lambda)=\{0\}$,}
\vspace{-2mm}
\item[(v)]{ $f(x)\in V_j^{(s)}(\Lambda)$ if and only if $ f(\theta^{-j}x)
\in V_{0}^{(s)}(\Lambda)$,}\vspace{-2mm}
\item[(vi)]{there exists a finite number of functions  $\phi_{\bar{n}}(x)\in V_0^{(s)}$, called
\underline{scaling} \underline{functions} such that
$\left\{\phi_{\bar{n}}(x-\lambda_k)\right\}_{k\in \bar{n},\, \bar{0} \leq \bar{n} \leq \bar{q}}$
is a Riesz   basis in $V_0^{(s)}$.}\vspace{-2mm}
\end{enumerate}
\end{prop}
The proof is straightforward from Definitions and Proposition \ref{prop1}.

\section{Spline wavelet basis}
\label{waveletConstruction}
As is well known, the whole wavelet basis is obtained by scaling the elements of a wavelet
subfamily living at a given scale of the multiresolution. For convenience, we choose
here the $0^{\rm th}$-scale and this subfamily
is precisely a basis of the orthogonal complement $W_{-1}^{(s)}(\Lambda)$ of
$V_{-1}^{(s)}(\Lambda)$ in $V_0^{(s)}(\Lambda)$:
\begin{equation}\label{rozklad}
V_0^{(s)}(\Lambda)=V_{-1}^{(s)}(\Lambda)\oplus_{\!\bot}
W_{-1}^{(s)}(\Lambda).
\end{equation}
This decomposition can be interpreted as the building up of the ``$(j=0)^{\rm th}$-scale''
content of the multiresolution ({\it i.e.} $V_0^{(s)} $) from the adding of
necessary ``details'' ({\it i.e.} $W_{-1}^{(s)}$)
to the existing content at
the next larger scale ({\it i.e.} $V_{-1}^{(s)} $).
A crucial step in the characterization of these ``details'' is the following existence theorem.
\begin{theom} (Bernuau \cite{bern1,bern2}
Let $\Lambda\in\R$ be a Delaunay set of finite local complexity, self-similar with
scaling factor $\theta > 1$. Then, for all $s\geq 2$, there exists a Riesz
basis of $W_{-1}^{(s)}(\Lambda)$ of the form
$$ \left\{\zeta_{\lambda_n}^{(s)}(x-\lambda_n),\ \lambda_{n}\in
\Lambda, \ \lambda_{n+1} \notin \theta \Lambda \right\},\quad {\rm where}\
  \left\{\zeta_{\lambda_n}^{(s)}(x) \right\}
$$
 is a finite set of functions with compact support.
\end{theom}

We here sketch the proof by just  listing a sequence of intermediate results given in \cite{bern1,bern2}. A preliminary characterization of
$W_{-1}^{(s)}$ is given by the following proposition. Let us recall that $H^s (\R)$ denotes the Sobolev space of
functions $f\in\LR$ such that $\cfrac[l]{\D^\alpha}{\D x^\alpha}
f$, with $\alpha\leq s$, is  element of $\LR$.
\begin{prop}
For $s\geq 2$, $ W_{-1}^{(s)} $ is the space of $s^{\text{th}}$ derivatives of elements in $
H^{2s}(\R)$ which vanish on $\theta\Lambda$:
 $$\ W_{-1}^{(s)}(\Lambda)
\equiv \cfrac[l]{\D^s}{\D x^s}K_{2s}=\left\{ \cfrac[l]{\D^s}{\D x^s} h, \ h \in H^{2s}(\R)\ \big| \
\forall \lambda\in\theta\Lambda, \ h(\lambda)=0 \right\}.
$$
\end{prop}
\begin{proof}
As indicated in  Equation (\ref{rozklad}) let us determine  the
orthogonal complement of $V_{-1}^{(s)}(\Lambda)$ in
$V_0^{(s)}(\Lambda).$ Let $f\in V_{-1}^{(s)}(\Lambda)$ and
$h=\cfrac[l]{\D^s}{\D x^s} h_1$, where $h_1\in K_{2s}.$ Then
$$\disty\int_{\R}\bar{f}h \, \D x = \left\langle f,\cfrac[l]{\D^s}{\D x^s} h_1\right\rangle = (-1)^s \left\langle
\cfrac[l]{\D^s}{\D x^s}f, h_1\right\rangle=0.$$

Conversely, let $f\in\LR$ and orthogonal to $W_{-1}^{(s)}(\Lambda) =
\cfrac[l]{\D^s}{\D x^s}K_{2s}$. Let us  choose a function $\varphi\in
C_0^{\infty}$ such that $ \varphi(0)=1$ and with such small compact support
that its translates $\varphi(x-\lambda), \lambda\in \theta\Lambda$, have disjoint
supports. For all $g\in C_0^{\infty}, \
g(x)-\sum_{\lambda\in\theta\Lambda}g(\lambda)\varphi(x-\lambda)$
is in $K_{2s}$. Accordingly $$\int_{\R}\bar{f}(x)\cfrac[l]{\D^s}{\D
x^s}\left(g(x)-
\sum_{\lambda\in\theta\Lambda}g(\lambda)\varphi(x-\lambda)\right)\D
x=0,$$ so we have $$\left\langle \cfrac[l]{\D^s}{\D x^s}
f,g\right\rangle = \sum_{\lambda\in\theta\Lambda}\left\langle f,
\cfrac[l]{\D^s}{\D x^s} \varphi(x-\lambda) \right\rangle g(\lambda),$$
which means that $f\in V_{-1}^{(s)}(\Lambda).$
\end{proof}
\begin{remark}
\! An equivalent characterization is to write that, for $s\geq 2,
W_{-1}^{(s)}(\Lambda)$ is the set of functions $f\in\LR, \
f=\cfrac[l]{\D^s}{\D x^s}h, $ for which $h\in V_0^{(2s)}{(\Lambda)}$
and $\left.h\right|_{\theta\Lambda}=0.$
\end{remark}
In view of this remark, the explicit construction of these wavelets necessitates the introduction of the following space
$$
\widetilde{W}_{-1}^{(s)}(\Lambda)=\left\{f\in
V_0^{(s)}(\Lambda)\, \big| \,  f{\mid_{\theta\Lambda}}=0 \right\}.
$$
This is a closed subspace
of $V_0^{(s)}({\Lambda})$. If we consider in particular
$\widetilde{W}_{-1}^{(2s)}(\Lambda)$, then, by  derivation
$\cfrac[l]{\D^s}{\D x^s}$ of its elements, we will obtain functions in $W_{-1}^{(s)}(\Lambda)$.
 Now, we need functions  in $\widetilde{W}_{-1}^{(s)}(\Lambda)$  with  minimal
 support. Let us consider the following subset of integers:
 \begin{equation}\label{setE}
 E=\left\{n\in\Z \, \mid \, \lambda_{n+1}\notin\theta\Lambda\right\}.
 \end{equation}
 Next, for all $n\in E$, let us define the unique function $\Psi_{n}^{(s)}\in
 \widetilde{W}_{-1}^{(s)}(\Lambda)$ satisfying the following
 conditions:
 \begin{enumerate}
 \item[(i)]{${\rm supp}\,\Psi_{n}^{(s)}$ is compact and included
 in  $[\lambda_{n},\infty)$,}
 \item[(ii)]{$\Psi_{n}^{(s)}(\lambda_{n+1})=1$,}
 \item[(iii)]{ $f\in  \widetilde{W}_{-1}^{(s)}(\Lambda),\
 {\rm supp}f\subset{\rm supp}\,\Psi_{n}^{(s)}\Rightarrow f\varpropto
 \Psi_{n}^{(s)}.$}
 \end{enumerate}
Then we have another important result.
 \begin{theom}
 The set $\left\{\Psi_{n}^{(s)},\ n\in E\right\}$ is a Riesz basis of the space
 $\widetilde{W}_{-1}^{(s)}(\Lambda)$.
 \end{theom}
 \noindent
The proof of this theorem is  also given in \cite{bern1,bern2} and goes through a list of intermediate properties which we give here without
proof:
\begin{proper}\label{support}
For all $n\in E$, the support of $\Psi_{n}^{(s)}$ is the interval
   $[\lambda_{n},\lambda_{n+N_{n}}]$, where $N_{n}$ is the smallest number equal or larger
   than $s$ such that \\
    $N_{n}=s+\#\{(\lambda_{n},\lambda_{n+N_{n}})\,\cap\,\theta\Lambda\}$.
\end{proper}
Note that $N_n$ is also defined as the smallest number such that  the interval
$(\lambda_{n},\lambda_{n+N_{n}})$ includes exactly $s-1$ points from $\Lambda\setminus\theta\Lambda.$
\begin{proper}\label{prop4}
There exists a finite set  $\mathcal{G}_{s}$ of functions with compact
support such that, for all $k\in E$, $\Psi_{k}^{(s)}(x+\lambda_{k})\in
\mathcal{G}_{s}$.
\end{proper}
\begin{proper}
All functions from $\widetilde{W}_{-1}^{(s)}(\Lambda)$ with  compact
support are  linear combination of the $\Psi_{k}^{(s)},\ k\in E$.
\end{proper}
\begin{proper}
The set of functions in $\widetilde{W}_{-1}^{(s)}(\Lambda)$ that have
compact support is dense in $\widetilde{W}_{-1}^{(s)}(\Lambda)$.
\end{proper}
\begin{proper}
For all $m\in\Z$, define $I_{m}$ as the subset of $k\in E$ for which
$\Psi_{k}^{(s)}$ is not  identically equal to zero on the interval
$[\lambda_{m},\lambda_{m+1}]$. Then $s-1\leq\#(I_{m})\leq s$.
\end{proper}
\begin{proper}
There are two constants $0<A\leq B$, depending on $s$ and
$\Lambda$ only, such that, for any sequence $(a_k)_{k\in E}$ with finite support, we have
 $$A\sum_{k\in E}|a_k|^2\leq \Big\Vert \sum_{k\in E} a_k
 \Psi_{k}^{(s)}\Big\Vert^2_2
 \leq B\sum_{k\in E}|a_k|^2.$$
\end{proper}
Note that if $A=1=B$ then the set $\left\{\Psi_{n}^{(s)},\ n\in E\right\}$ is an orthonormal basis.
The main result of this section is a consequence of all the above statements:
\begin{theom}\label{main_theorem} (Bernuau \cite{bern1,bern2}) 
Let $\Lambda = \left\{\lambda_n   \right\}_{n\in \Z} \subset \R$ be a Delaunay set of finite local complexity, self-similar
with factor $\theta>1.$ Let us denote the elements of $\theta^{-1}\Lambda$ by $\kappa_n, n\in \Z$, $\kappa_n = \theta^{-1}\lambda_n$. Then for all $s>1$ there exists a Riesz basis
of $\LR$ of the form:
\begin{equation}
\left\{\theta^{j/2}\psi_{\kappa_n}^{(s)}(\theta^{j}x-\kappa_n),\
\kappa_n \in  \theta^{-1} \Lambda, \ \kappa_{n+1} \notin \Lambda,\ j\in\Z \right\},
\end{equation}
where $\left\{\psi_{\kappa_n}^{(s)}, \right\}$ is a finite set
of compactly supported functions of order $C^{s-2}$.
\end{theom}
\begin{proof}
For $\lambda_n \in \Lambda$, define $\zeta_{\lambda_n}^{(s)}(x-\lambda_n)=\cfrac[l]{\D^s}{\D
x^s}\Psi_{n}^{(2s)}(x).$ Then, according to  Property
\ref{prop4}, $\left\{ \zeta_{\lambda_n}^{(s)}\right\}$ is a finite set for
$\lambda_{n+1}\notin \theta\Lambda$. These functions together with all their
admissible translates $\zeta_{\lambda_n}^{(s)}(x-\lambda_n)$ form a Riesz basis of $W^{(s)}_{-1}(\Lambda)$. We know that
the sequence of spaces $\left( V_j^{(s)}(\Lambda) \right)_{j\in\Z}$ is
a multiresolution analysis of $\LR$, so we get the orthogonal sum
$$
\LR=\bigoplus_{j \in \Z}\,\Bot W_j^{(s)}(\Lambda)
$$
 where $f(x)\in
W_j^{(s)}(\Lambda)\Longleftrightarrow f(\theta^{-j}x)\in
W_0^{(s)}(\Lambda).$

For convenience we shift by one the scale and choose  the $L^2$-normalization
in order to define
\begin{equation}\label{psi}
\psi_{\kappa_n}( x) = \frac{\zeta_{\lambda_n} (\theta x)}{\Vert \zeta_{\lambda_n} (\theta x) \Vert} \in W_0^{(s)}(\Lambda).
\end{equation}
Then we can assert that the set
$\left\{\theta^{j/2}\psi_{\kappa_n}^{(s)}(\theta^{j}x-\kappa_n),\
\kappa_n \in  \theta^{-1} \Lambda, \ \kappa_{n+1} \notin \Lambda\right\}$ is a Riesz basis
of $W_j^{(s)}(\Lambda)$. By union on all $j\in\Z$ we obtain the
theorem.
\end{proof}

\section{Self-similar discretizations of $\pmb{\R}$ with UPV$\bmth{ _2}$ scaling factor}

\subsection{Two-letter substitution sequences, quadratic PV numbers, and beta-integers }
Let $a$ be a positive integer  and  consider the following
two types of two-letter substitution  ($\L$ for ``long'' and $\S$ for ``short'')
\begin{equation}\label{substi-}
 \varsigma : \left\{
  \begin{array}{rcl}
    \L & \rightarrow & \overset{a\ \text{times}}{{\overbrace{\L \cdots\L}}}\S,\\
    \S & \rightarrow & \L,
  \end{array} \ \mbox{with} \ a \geq 1. \right.\vspace{-2mm}
\end{equation}

\begin{equation}\label{substi+}
 \varsigma : \left\{
  \begin{array}{rcl}
    \L & \rightarrow & \overset{a-1\ \text{times}}{{\overbrace{\L \cdots\L}}}\S,\\
    \S & \rightarrow & \overset{a-2\ \text{times}}{{\overbrace{\L \cdots\L}}}\S,
  \end{array} \ \mbox{with} \ a \geq 3. \right.\vspace{-2mm}
\end{equation}
The corresponding substitution matrices read respectively $\bigl( \begin{smallmatrix} a & 1  \\
    1 & 0 \end{smallmatrix} \bigr)$  and $\bigl( \begin{smallmatrix} a-1 & 1  \\
    a-2 & 1 \end{smallmatrix} \bigr)$. The characteristic equation for the former is
 \begin{equation}\label{quad-}
X^2 - aX -1 = 0 \ \mbox{with roots} \ \left\{  \begin{array}{l}\frac{a + \sqrt{a^2 + 4}}{2} \equiv \beta, \\
\frac{a - \sqrt{a^2 + 4}}{2} = -\frac{1}{\beta} \equiv \beta' = a - \beta.
\end{array} \right.
\end{equation} For the second matrix we have
\begin{equation}\label{quad+}
X^2 - aX +1 = 0 \ \mbox{with roots} \  \left\{  \begin{array}{l} \frac{a + \sqrt{a^2 -  4}}{2} \equiv \beta, \\
\frac{a - \sqrt{a^2 - 4}}{2}= \frac{1}{\beta} \equiv \beta' = a - \beta.\end{array} \right.
\end{equation}
In both cases, the largest root $\beta$ is  $ >1$ whilst its {\it
Galois conjugate} $\beta' \in (-1,1)$. For this reason, one says
that $\beta$ is a quadratic  {\it Pisot-Vijayaraghavan} and a unit
for being invertible in their respective extension ring $\Z
\lbrack \beta \rbrack $ (the lowest-degree coefficient in
\eqref{quad-}  and \eqref{quad+} is $\pm 1 $). Henceforth, we
shall denote by $\BPM$ (resp. $\BPP$) the set of UPV$ _2$'s
obeying \eqref{quad-} (resp. \eqref{quad+}).

   Let
us now associate to the letter $\L$ a tile of length $1$ and to $\S$  a tile of length
 $1/\beta$ for the case \eqref{substi-} and   $1 - 1/\beta$ for the case \eqref{substi+}.
Starting from the origin of the real line with
$\L$ on the
 right, we apply to it the substitution
 $\varsigma^{\infty}$.  The set
 of  nodes of the resulting tiling of the half-line is called the set of positive
``beta-integers'' and is denoted by $\ZB^+$. Taking the symmetric $- \ZB^+$ of
$\ZB^+$ with respect of the origin, we obtain the set of ``beta-integers'' $\ZB = \ZB^+ \cup
(-\ZB^+) $. This set is clearly a Delaunay set which is of finite type complexity and self-similar
with factor $\beta$: $\beta \ZB \subset \ZB $. By construction,
it is also symmetrical with respect to the origin: $-\ZB = \ZB $.
This name of {\it beta-integers} comes from the fact that if we
retain in the set of real numbers all those numbers which
are polynomial in $\beta$ when written in ``basis $\beta$'', then
they form a totally ordered discrete set that precisely coincides with  $\ZB$. We recall
here that the writing of a real positive number $x$ in irrational basis $\beta > 1$ means a unique
$\beta$-expansion $ x = \sum_{l = -\infty}^j x_l \beta^l \equiv x_j x_{j-1} \cdots x_1 x_0 \cdot
  x_{-1} \cdots x_{l} \cdots $ in which the expansion coefficients assume their values in the
alphabet $\{0, 1, \dotsc , \lfloor \beta \rfloor \ \mbox{integer part of}\ \beta\, \}$ in agreement
with the existence of allowed  words determined by the so-called {\it greedy algorithm} (for more
precisions, see for instance  \cite{gaz1, gaz2}). In the ``minimal'' case $a=1$ of the first category
\eqref{quad-},
$\beta = \tau = (1+ \sqrt5)/2 \approx 1.618\cdots$, the rules are
particularly simple, since there the alphabet is $\{ 0, 1 \}$ and the constraint  is that
no two-letter string $11$ should appear in any $\tau$-expansion words.  The first pieces around the
origin  of the associated tiling  are shown in the picture below.

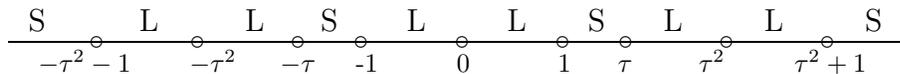
\begin{figure}[h]
\begin{center}
\unitlength=.67cm
\begin{picture}(17,3)\label{picture}
\put(0,2){\line(1,0){17.85}} \put(9,2){\circle{.2}}
\put(7,2){\circle{.2}} \put(5.75,2){\circle{.2}}
\put(3.75,2){\circle{.2}} \put(1.75,2){\circle{.2}}
\put(11,2){\circle{.2}} \put(12.25,2){\circle{.2}}
\put(14.25,2){\circle{.2}} \put(16.25,2){\circle{.2}}
\put(8.9,1.4){\footnotesize 0} \put(6.9,1.4){\footnotesize -1}
\put(5.4,1.4){\footnotesize$-\tau$} \put(3.6,1.4){\footnotesize
$-\tau^2$} \put(0.6,1.4){\footnotesize$-\tau^2 -1$}
\put(10.9,1.4){\footnotesize 1}
\put(12.1,1.4){\footnotesize$\tau$}
\put(13.7,1.4){\footnotesize$\tau^2$}
\put(15.6,1.4){\footnotesize$\tau^2 + 1$} \put(9.9,2.2){L}
\put(11.5,2.2){S} \put(13,2.2){L} \put(15,2.2){L} \put(17,2.2){S}
\put(7.9,2.2){L} \put(6.2,2.2){S} \put(4.7,2.2){L} \put(2.6,2.2){L}
\put(0.4,2.2){S}
\end{picture}
\end{center}
\vspace*{-14mm}
 \caption{Tau-integers around the point 0.}
\end{figure}
Note that the case coming just next this minimal one, namely $a=2$, is the ``octogonal''
number $\beta = \omega = 1 +\sqrt2$.

The minimal case in the second category corresponds to $a=3$ and yields the UPV$ _2$
$\beta = (3 + \sqrt5)/2 = \tau^2 = 1 + \tau \approx 2.618\cdots$. The alphabet
is now $\{ 0, 1, 2\}$ and the constraint  is that
no two-letter string $2\overset{n\,\text{times}}{\overbrace{1\cdots 1}}2$, $n\geq 0$,
 should appear in any  $\tau^2$-expansion words.
The first pieces
of the associated tiling on the right of the origin   are shown in the picture below.

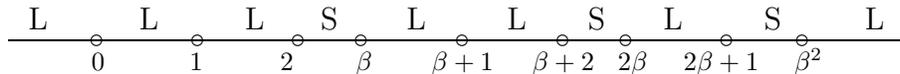
\begin{figure}[!h]
\begin{center}
\unitlength=.67cm
\begin{picture}(17,2.5)
\put(0,2){\line(1,0){17.85}}
\put(9,2){\circle{.2}}
\put(7,2){\circle{.2}}
\put(5.75,2){\circle{.2}}
\put(3.75,2){\circle{.2}}
\put(1.75,2){\circle{.2}}
\put(11,2){\circle{.2}}
\put(12.25,2){\circle{.2}}
\put(14.25,2){\circle{.2}}
\put(15.75,2){\circle{.2}}
\put(8.4,1.4){\footnotesize $\beta + 1$}
\put(6.9,1.4){\footnotesize $\beta$}
\put(5.4,1.4){\footnotesize$2$}
\put(3.6,1.4){\footnotesize$1$}
\put(1.65,1.4){\footnotesize$0$}
\put(10.4,1.4){\footnotesize$\beta + 2$ }
\put(12.1,1.4){\footnotesize$2\beta $}
\put(13.4,1.4){\footnotesize$2\beta + 1$}
\put(15.6,1.4){\footnotesize$\beta^2$}
\put(9.9,2.2){L}
\put(11.5,2.2){S}
\put(13,2.2){L}
\put(15,2.2){S}
\put(17,2.2){L}
\put(7.9,2.2){L}
\put(6.2,2.2){S}
\put(4.7,2.2){L}
\put(2.6,2.2){L}
\put(0.4,2.2){L}
\end{picture}
\end{center}
\vspace*{-14mm}
 \caption{Beta-integers around the point 0 for $\beta = \tau^2$.}
\end{figure}

\subsection{Model set discretizations of $\pmb{\R}$}
Another way of obtaining self-similar  Delaunay sets with finite local complexity is
to resort to the so-called Cut-and-Project method which has become
like a paradigm in quasicrystalline studies. We shall adopt here the
formalism set up by Meyer \cite{Me95,Mo2}.
A 1+1-{\em cut and project scheme} is the following
\begin{equation}\label{cutpro}
\begin{array}{ccccc}
\R & \stackrel{\pi_1}{\longleftarrow}
 \R  \times  \R \stackrel{\pi_2}{\longrightarrow} & \R \\
&  \cup  & \\
&  D  &
\end{array}
\end{equation}
where  $D$ is a lattice.
The projection $\pi_{1}|_{D}$ is 1-to-1, and $\pi_2(D)$ is dense in $\R$.

Let $M=\pi_1(D)$ and set $ ^\ast=\pi_2 \circ (\pi_{1}|_{D})^{-1}$
\begin{equation}\label{ast}
 ^\ast : M \longrightarrow \R .
\end{equation}

The set
$\Lambda \subset \R$ is a {\em model set}
if there exist a cut and project scheme and a relatively compact set
$\Omega \subset \R$ of non-empty interior such that
\begin{equation}\label{mod}
\Lambda = \{x \in M \mid x^\ast \in \Omega\} \equiv {\Sigma}^{\Omega}.
\end{equation}
The set $\Omega$ is called a {\em window}.

As an illustration, we describe one type of Fibonacci chain with scaling factor
$\beta = \tau^2$.
Consider the  cut and project scheme \eqref{cutpro} with $D=\Z^2$ and
 $\pi_{1}(\Z^2) \sim \Z[\tau]=\{a+b \tau \mid a,b \in \Z\}$. The map \eqref{ast}
is identical up to a factor to the Galois ring automorphism
$x = a + b\tau \longrightarrow x' = a - \frac{b}{\tau} $.
An example of  Fibonacci chain $\FT$ \cite{Mopa} is given by choosing the
semi-open interval $\lbrack0,1) $ as a window.
\slacs{.6mm}
\begin{eqnarray*}
{\FT}&=&\{x=a+b \tau \mid x'=a-\frac{b}{\tau} \in \Omega=[0,1)\}\\
 & = & \{ \ldots, -\tau^3, -\tau,0,\tau^2,\tau^3 +1,\tau^4, \ldots\}.
\end{eqnarray*}\slacs{1.2mm}
It is the set of left endpoints of a
quasiperiodic tiling of $\R$ with 2 tiles L et S,
of respective length $\tau^2$ and $\tau$,
generated by the substitution rules
\begin{equation}\label{subfib}
 \varsigma : \left\{
\begin{array}{ccl}
\L &  \rightarrow & \L\L\S,\\ \S & \rightarrow & \L\S.
\end{array}\right.  \vspace{-2mm}
\end{equation}
Starting from S to the left  and from L to the right we get a biinfinite word
$$
\cdots \text{LLSLS}\mid  \text{LLSLLSLS} \cdots .
$$
Note that
$\tau^2 \FT  \subset \FT$
and that this tiling is, by construction, {\em stone inflation}, which means that
all the tiles when scaled by the factor $\tau^2$ can be packed face-to-face from the original ones.
Therefore, the UPV$ _2$ $\tau^{2} \in \BPP$ is the scaling factor of the substitution $\varsigma$.
The first pieces of this Fibonacci tiling around the origin are shown in the picture below.

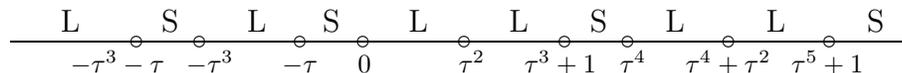
\begin{figure}[h]
\begin{center}
\unitlength=.67cm
\begin{picture}(17,3)
\put(0,2){\line(1,0){17.85}}
\put(9,2){\circle{.2}}
\put(7,2){\circle{.2}}
\put(5.75,2){\circle{.2}}
\put(3.75,2){\circle{.2}}
\put(2.50,2){\circle{.2}}
\put(11,2){\circle{.2}}
\put(12.25,2){\circle{.2}}
\put(14.25,2){\circle{.2}}
\put(16.25,2){\circle{.2}}
\put(8.9,1.4){\footnotesize$ \tau^2$}
\put(6.9,1.4){\footnotesize 0}
\put(5.4,1.4){\footnotesize$-\tau$}
\put(3.5,1.4){\footnotesize$-\tau^3$}
\put(1.2,1.4){\footnotesize$-\tau^3 -\tau$}
\put(10.2,1.4){\footnotesize $\tau^3 + 1$}
\put(12.1,1.4){\footnotesize$\tau^4$}
\put(13.4,1.4){\footnotesize$\tau^4 + \tau^2$}
\put(15.5,1.4){\footnotesize$\tau^5 + 1$}
\put(9.9,2.2){L}
\put(11.5,2.2){S}
\put(13,2.2){L}
\put(15,2.2){L}
\put(17,2.2){S}
\put(7.9,2.2){L}
\put(6.2,2.2){S}
\put(4.7,2.2){L}
\put(3,2.2){S}
\put(1,2.2){L}
\end{picture}
\end{center}
\vspace*{-14mm}
 \caption{Fibonacci chain around the point 0.}
\label{picture1}
\end{figure}

More generally, for $\beta \in \BP$ we shall consider the model set:
\begin{equation}
\FB \equiv \{x=\Z\lbrack \beta \rbrack\, \mid\,  x' \in \Omega= \lbrack 0,1)\}
\end{equation}
where the prime `` ' '' designates the  ring automorphism
$x = a + b\beta \longrightarrow x' = a + b\beta' $.
Let us give a set of properties concerning $\FB$ when $\beta \in \BP$. Proofs can be found in
\cite{BFGK,BFGK2,mapape}.

\begin{prop}
Suppose that $\beta \in \BPM$. Then the model set $\FB$ is a self-similar  Delaunay set of finite local complexity,
with scaling factor $\beta^2$. It can be  characterized either as a subset of the beta-integers or
as set of nodes of the stone inflation tiling associated to a substitution sequence. More precisely:
\begin{enumerate}\vspace{-2mm}\label{FB-}
\item[(i)] $\FB$ is obtained from  $\ZB$ through the sieving procedure
\begin{equation}
\FB = \{ x\in \ZB \, \mid \, x' \in \lbrack 0,1) \}.
\end{equation}
\item[(ii)] $\FB$ is the set of left endpoints of a
quasiperiodic tiling of $\R$ with 2 tiles $\L$ et $\S$,
of respective length $\beta + 1$ and $\beta$,
generated by the substitution rules
\begin{equation}\label{subeta-}
 \varsigma : \left\{
\begin{array}{rcl}
    \L & \rightarrow &\L\overset{a\ \text{times} }{\overbrace{ \overset{a-1\ times}{\overbrace{\S \cdots\S}}\L
\overset{a-1\ times}{\overbrace{\S \cdots\S}}\L\cdots \overset{a-1\ times}{\overbrace{\S \cdots\S}}\L}}
\overset{a\ times}{\overbrace{\S \cdots\S}},\\
    \S & \rightarrow & \L\overset{a-1\ times }{\overbrace{ \overset{a-1\ times}{\overbrace{\S \cdots\S}}\L
\overset{a-1\ times}{\overbrace{\S \cdots\S}}\L\cdots \overset{a-1\ times}{\overbrace{\S \cdots\S}}\L}}
\overset{a\ times}{\overbrace{\S \cdots\S}},
  \end{array} \right.
\end{equation}
with $ a \geq 1$.
The tiling is obtained by starting from $\S$-origin-$\L$ in both directions.
\end{enumerate}
\end{prop}

\begin{prop}
Suppose that $\beta \in \BPP$. Then the model set $\FB$ is a self-similar  Delaunay set of finite local complexity,
with scaling factor $\beta$. It can be  characterized either as a subset of {\it decorated} beta-integers or
as set of nodes of the stone inflation tiling associated to a substitution sequence. More precisely:
\begin{enumerate}\vspace{-2mm}\label{FB+}
\item[(i)] $\FB$ is obtained from the decorated beta-integers
 $\widetilde{\Z}_\beta  \stackrel{def}{=} \ZB + \{ 0, \pm \frac{1}{\beta}\}$ through the sieving
procedure
\begin{equation}
\FB = \{ x\in \widetilde\Z_\beta \, \mid \, x' \in \lbrack 0,1) \}.
\end{equation}
\item[(ii)] Alternatively, $\FB$ reads as
\begin{equation}\label{fibco}
\FB = \beta \left\lbrack \ZB^+ \cup \left((-\ZB^+)\setminus\{0\} + \frac{1}{\beta} \right)\right\rbrack.
\end{equation}
\item[(iii)] $\FB$ is the set of left endpoints of the
quasiperiodic tiling of $\R$ with 2 tiles $\L$ et $\S$,
of respective length $\beta$ and $\beta - 1$,
generated by the substitution rules \eqref{substi+}.
The tiling is obtained by starting from $\S$-origin-$\L$ in both directions.
\end{enumerate}
\end{prop}
Note that the minimal case $ \beta = \tau$ is exceptional in the sense that
there is equality between both extension rings $\Z \lbrack \tau  \rbrack = \Z \lbrack \tau^2  \rbrack $
and so one can deduces from \eqref{fibco}:
$\FT = \beta \left\lbrack \ZB^+ \cup \left((-\ZB^+)\setminus\{0\} + \frac{1}{\beta} \right)\right\rbrack $, with $\beta = \tau^2 $.

 \section{Haar wavelets for  $\bmth{\beta}$-integers}
 We now turn our attention on the construction  of multiresolution analysis
of the Haar type based on non-negative $\beta$-integers
$\Z_\beta^+$, the extension to the full $\Z_\beta$ being carried out by
simple symmetry with respect to the origin. Let us denote by  space $V_0^+$
 the closure in $L^2(\R^+)$ of the linear span of all positive admissible translates of
normalized characteristic functions
$\phi_\L(x)$ and
$\phi_\S(x)$ supported by intervals
 of lengths $|\L|$ and $|\S|$ respectively. More precisely,
 $$
 V_0^+=
\overline{\mbox{span}\left\{\phi_\L(x-\lambda_\L),\phi_\S(x-\lambda_\S)
 \right\}_{\lambda_\L\in\Lambda_\L^+,\lambda_\S\in\Lambda_\S^+}},
 $$
 where $\Lambda_\L^+$ is the set of left-hand ends of tiles
 L in $\ZB^+$ and $\Lambda_\S^+$ is the set of left-hand ends of tile
 S in $\ZB^+$.

Two possibilities exist for constructing basis made up with piecewise constant functions. Either one builds  orthonormal basis,
at the price of increasing the number of {\it mother} functions,
or one just requires   a Riesz basis and then we need  two {\it mother} functions only.

\subsection{Orthogonal basis of Haar type}

 Let us first  discuss the  case (\ref{substi-}).
 Due to the two-letter substitution rules, two possible distances between
points exist in the associated tiling,
namely $|\L|=1$
 and $|\S|=1/\beta$. Consistently, two scaling functions exist,
 one per type of tile, and their refinement
 equations are precisely based on these substitution rules.
 For $a\geq 1$ the two scaling functions read:
 \slacs{.6mm}
\begin{align}
 \phi_\L(x)&=\mathbf{1}_{[0,1)}(x),   \\
  \phi_\S(x-a) &=\beta^{1/2} \mathbf{1}_{[a,a+1/\beta)}(x).
\end{align}

 Consequently, the scaling (or refinement) equations for $a\geq 2$ are the
following:
\begin{align}
 \phi_\L(x) &=\sum_{l=0}^{a-1}\phi_\L(\beta x-l)+
  \beta^{-1/2}\phi_\S(\beta x-a),\\
  \phi_\S(x-a)& = \beta^{1/2} \phi_\L(\beta x-a \beta) .
\end{align}

The corresponding orthonormal wavelet basis is built from $a$ mother
wavelets
$\{\psi_{\L,i}\}_{i=0}^{a-1}$.
  One method of construction of these functions is to Gram-Schmidt
orthogonalize and normalize
  in the sense of $L^2$-norm the set of functions
$\{\phi_\L(x),\phi_\L(\beta x),$
$\phi_\L(\beta x-1),
  \dots,\phi_\L(\beta x-(a-1))\}$ living on the interval $[0,1]$.
  An orthonormal basis for $V_1^+=V_0^+\oplus_{\!\bot} W_0^+$ is then obtained by
collecting together all
translates of our orthonormal set at points of
  $\Lambda_\L^+$ and all translates of the $\phi_\S$'s  at points of
$\Lambda_\S^+$.
  After removing functions in $V_0^+$ we get the basis of $W_0^+$.

 The case $a=1$, $\beta=\tau$ is particularly interesting. In this case, $\Lambda_\L^+ = \tau \Z_{\tau}^+$ and $\Lambda_\S^+ = \tau^2
\Z_{\tau}^+ + 1$. There are two scaling functions but only one wavelet, and so the
refinement equations read as:
\begin{align}
 \phi_\L(x)&=\phi_\L(\tau x)+
  \tau^{-1/2}\phi_\S(\tau x-1),\nonumber \\
  \phi_\S(x-1) &=  \tau^{1/2} \phi_\L(\tau x- \tau),\nonumber\\
  \psi_{\L}(x)&= \tau^{-1/2} \phi_\L(\tau x)-\phi_\S(\tau x-1).
 \label{Ztau-wavelets}
\end{align}
Note that in the present case the following conditions are equivalent:
\begin{equation}
\lambda_n \in \tau^{-1}\Z_{\tau}^+ \ \text{and} \ \lambda_{n+1} \notin \Z_{\tau}^+ \, \Leftrightarrow \, \lambda_n \in
\tau^{-1}\Lambda_{\L\S}^+ \, \Leftrightarrow \, \lambda_n \in \tau\Z_{\tau}^+ \equiv \Lambda_\L^+,
\end{equation}
where we have denoted by $\lambda $ the generic elements of $\tau^{-1} \Z_{\tau}^+$.
Consequently, the orthonormal basis of $V_1^+$ is given by:
\begin{equation}
\left\{\phi_\S(x - \lambda)  \right\}_{\lambda \in (\tau^2 \Z_{\tau}^+ + 1)}\cup \left\{\phi_\L(x - \lambda)  \right\}_{\lambda \in
\tau\Z_{\tau}^+}\cup \left\{\psi_\L(x - \lambda)  \right\}_{\lambda \in \tau \Z_{\tau}^+},
\end{equation}
 and the orthonormal (``tau'') Haar basis of $L^2(\R^+)$ is the set
\begin{equation}
\left\{\tau^{j/2}\psi_\L(\tau^j x - \lambda)  \right\}_{j \in \Z,\lambda \in
\tau \Z_{\tau}^+} .
\end{equation}

The construction of the Haar wavelet basis corresponding to the
substitution of the second type is carried out in
a similar way. Lengths of tiles are respectively
$|\L|=1$ and $|\S|=1-1/\beta$.   For $a\geq 3$ we have for the two scaling
functions:
\begin{align}
 \phi_\L(x) &= \mathbf{1}_{[0,1)}(x),   \\
 \phi_\S(x-a+1)& = (1-1/\beta)^{-1/2} \mathbf{1}_{[a-1,a-1/\beta)}(x).
\end{align}
Consequently, the refinement equations now read:
\begin{align}
 \phi_\L(x)&=\sum_{l=0}^{a-2}\phi_\L(\beta x -l)+
  (1-1/\beta )^{1/2}\phi_\S(\beta x-(a-1)), \\
  \phi_\S(x-a+1)&= (1-1/\beta )^{-1/2} \sum_{l=0}^{a-3}\phi_\L(\beta x - \beta(a-1) - l) +\nonumber \\
&\phantom{=} +
  \phi_\S(\beta x-(a-1)\beta-a+2).
 \end{align}

 The corresponding orthonormal wavelet basis  is constructed from the
mother wavelet sets
$\{\psi_{\L,i}(x)\}_{i=0}^{a-2}$ and
   $\{\psi_{\S,i}(x)\}_{i=0}^{a-3}$.
Following the same method as in the previous case,  we
first pick  the functions $\{\phi_\L(x),\phi_\L(\beta
x),\phi_\L(\beta x-1),
  \dots,\phi_\L(\beta x-(a-1))\}$ living on the interval $[0,1]$
  and $\{\phi_\S(x),\phi_\L(\beta x),\phi_\L(\beta x-1),
  \dots,\phi_\L(\beta x-(a-2))\}$ living on  $[0,1-1/\beta]$. We
then
proceed to the Gram-Schmidt orthogonalization and $L^2$-normalization.
The orthonormal basis of $V_1^+$ is obtained by collecting together all
translates of the first orthonormal set at points of
  $\Lambda_\L^+$ and all translates of the second orthonormal set  at points
of
$\Lambda_\S^+$.
  After removing functions in $V_0^+$ we get the basis of $W_0^+$.

For the negative part of $\Z_\beta$ we just have to mirror the supports of
scaling functions and wavelets.

\subsection{Non-orthogonal basis of Haar type}
We still have  same scaling functions as in the orthogonal case. This means that
all spaces
$V_j^+$ are identical to the previous ones, but the basis of the orthogonal
complement $W_0^+$ of $V_0^+$ in $V_1^+$ is
just required to be normalized.
But the case $\beta=\tau$ which leads to the same result as in the above,
 there are generically  two mother wavelets
$\psi_{\L\L}$ and $\psi_{\L\S}$ which live on tiles $\beta^{-1}\L\L$,
resp. $\beta^{-1}\L\S$. The orthogonal complement
of $V_0^+$ in $V_1^+$ reads:
$$
W_0^+=\overline{\mbox{span} \left\{\psi_{\L\L}(x-\lambda_{\L\L}),
\psi_{\L\S}(x-\lambda_{\L\S})\right\}_{\lambda_{\L\L}\in\beta^{-1}\Lambda_{\L\L}^+,\lambda_{\L\S}\in\beta^{-1}\Lambda_{\L\S}^+}},
$$
where $\Lambda_{\L\L}^+$ (resp. $\Lambda_{\L\S}^+$) is the set of
left-hand ends of  words LL (resp. LS) in $\ZB^+$.

In the substitutional case (\ref{substi-}) the refinement
 equations for wavelets are:
\begin{align*}
  \psi_{\L\L}(x)&= \left(\frac{\beta }{2}\right)^{1/2}( \phi_\L(\beta x)-\phi_\L(\beta
x-1)),\\
   \psi_{\L\S}(x-(a-1)/\beta) &= 
\left(\frac{\beta }{\beta + 1}\right)^{1/2}(  \phi_\L(\beta
x-(a-1))-\beta^{1/2}\phi_\S(\beta x-a)).
\end{align*}

The construction of the Haar wavelet basis corresponding to the
substitution of the second type is carried out in a similar way.
The corresponding wavelets are given by:
\begin{align*}
  \psi_{\L\L}(x)&= \left(\frac{\beta }{2}\right)^{1/2}( \phi_\L(\beta x)-\phi_\L(\beta
x-1)),\\
   \psi_{\L\S}(x-(a-2) /\beta) &= \nonumber\\
 \left(\frac{(a-1)\beta -1}{2\beta - 1}\right)^{1/2}&( \phi_\L(\beta
x-a+2)-(1-1/\beta)^{-1/2}\phi_\S(\beta x-a+1)).
  \end{align*}
  For the negative part of $\Z_\beta$
scaling functions and their equations are still the same but instead of
the wavelet $\psi_{\L\S}$ we now have to deal with a wavelet of
the type $\psi_{\S\L}$.

\section{Lexicographical analysis of the Fibonacci chain}
We now focus on the rescaled version $ \Lambda \stackrel{\rm def}{=}\FT/\tau^2 $ of the Fibonacci chain $\FT$.
This is a standard example of model set, well known by quasicrystallographers, and most of the obtained results about it are
easily extendable to other cases described in Section 4. With the notation
\eqref{mod} this model set reads
$\Lambda=
\Sigma^{[0,\tau^{2})}$. We   also view $\Lambda \equiv\left\{ \lambda_n\right\}_{n\in\Z}$ as the increasing sequence  of  points
$\lambda_n<\lambda_{n+1}$ with $\lambda_0=0.$
  Let us give  a list of important properties stemming from the
interval nature of the window:
 \begin{itemize}
 \item {The model set $\Lambda$ is self-similar with factor $\tau^2$:$\ \tau^2 \Lambda
 \subset \Lambda$. More precisely, it is invariant under the affine-linear actions:
 \begin{equation}
 \label{afli}
 \tau^{2}{\Lambda}={\Sigma}^{[0,1)}\subset{\Lambda},\quad
 \tau^{2}{\Lambda}+{\Sigma}^{[0,\tau]}={\Lambda}.
\end{equation} }
\item{For a given $n$, $\Lambda$ reads as a partition into subsets of $n$-letter
words $\Lambda_{a_1 a_2 \dots a_n}=\left\{ \lambda\,|\lambda
\mbox{ is the left-hand end of the $n$-letter word } a_1 a_2 \dots a_n\right.$,
$\left. a_i\in\{\L,\S\}\right\}.$} More precisely, we  have the following tree-like
{\it Fibonacci} hierarchy of partitions:
\begin{eqnarray*}
 \Lambda=&\Lambda_\L&\cup\qquad\Lambda_\S, \\
\Lambda=&\overbrace{\Lambda_{\L\L}\cup\Lambda_{\L\S}}&\cup\qquad\Lambda_{\S\L}\\
\Lambda=&\Lambda_{\L\L\S}\cup\Lambda_{\L\S\L}&\cup\overbrace{\Lambda_{\S\L\L}\cup\Lambda_{\S\L\S}}
\\
  &\vdots&
\end{eqnarray*}
There are only $n+1$ words built up from $n$ letters (the
substitution is Sturmian with {\em minimal complexity}).
\item{The above sequence of partitions is in one-to-one correspondence with
partitions of the interval window $[0,\tau^2)$. Accordingly,
each element in the partition sequence is also a model set:
 $\Lambda_{a_1 a_2 \dots a_n}=
{\Sigma}^{[\omega_1,\omega_2)},$} where
$\omega_1,\omega_2\in\Z[\tau].$
\end{itemize}

\begin{figure}
\begin{center}
\includegraphics[angle=0,width=6.5cm]{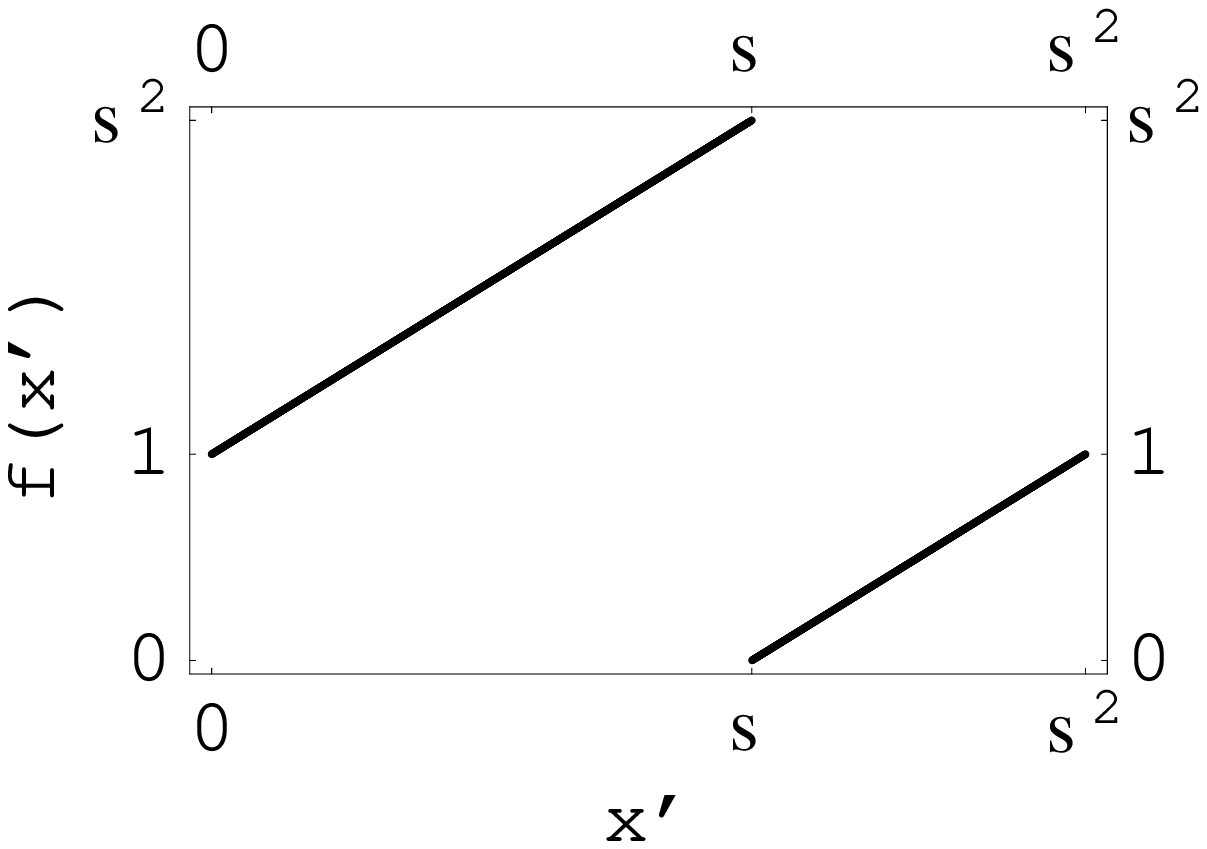}
\includegraphics[angle=0,width=7cm]{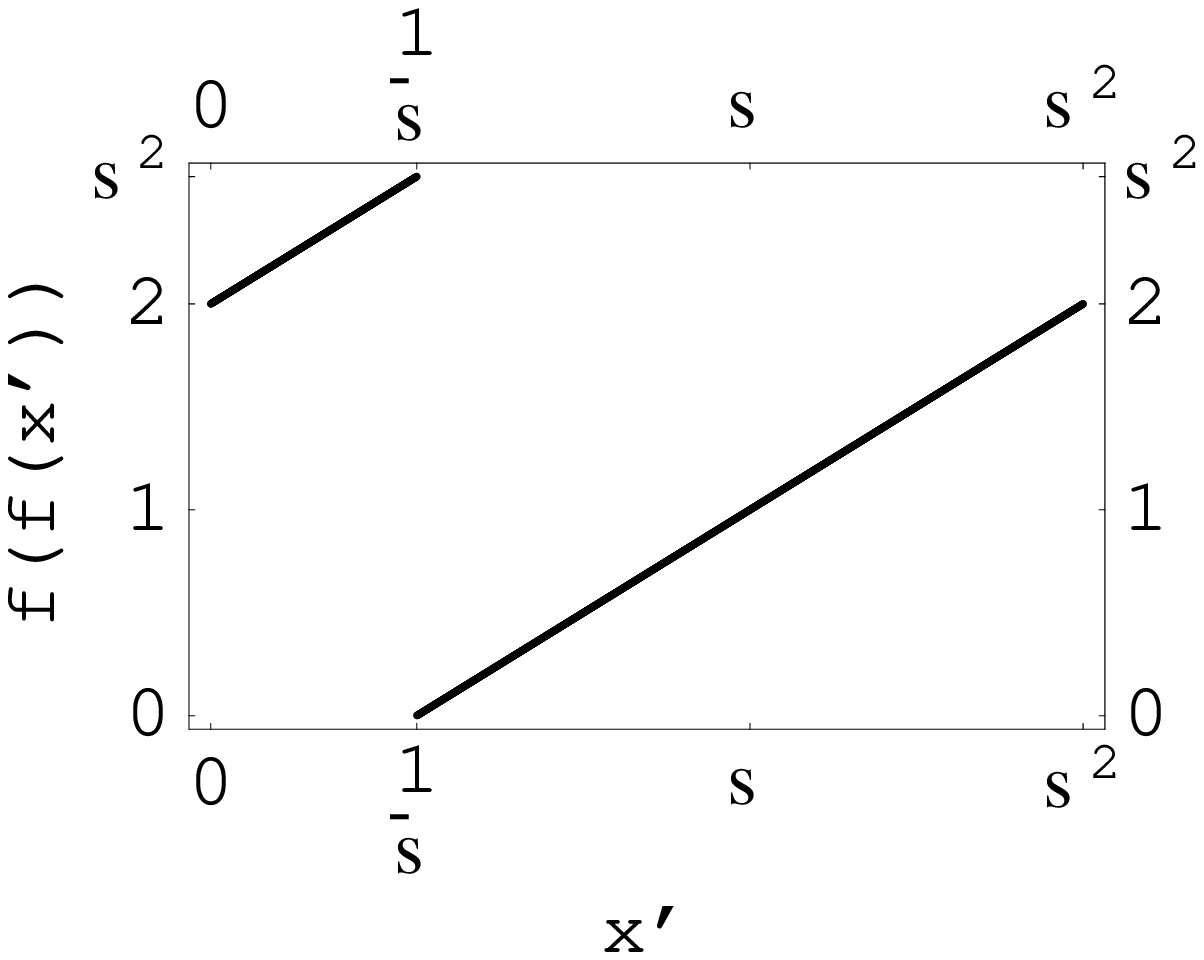}
\vspace*{-1cm}
\end{center}
\caption{Conjugate right nearest neighbour function $f$ (left graph) and its iterated $f\circ f $ (right graph).}
\label{function_f}
\end{figure}

The properties of sets $\Lambda_{a_1 a_2 \dots a_n}$,
are suitably encoded by the one-to-one map $f:\Lambda' \longrightarrow \Lambda'$
introduced in \cite{mapape} and defined by $f(\lambda_n')=\lambda_{n+1}'$ where
$\lambda_n\in\Lambda$. This function uniquely determines
the nearest right neighbour of any point  $\lambda\in\Lambda$. Its reciprocal
$f^{-1}(\lambda_n')=\lambda_{n-1}'$  corresponds
to the nearest left neighbour. The function $f$ is explicitly given by:
\begin{equation}
f(x')=
  \begin{cases}
   x'+1 & \textrm{if } x\in\Lambda_\L \ ( \textrm{since }\lambda_{n+1}=\lambda_n+1), \\
    x'-\tau & \textrm{if } x\in\Lambda_\S \ (\textrm{since } \lambda_{n+1}=\lambda_n+1/\tau).
  \end{cases}
\end{equation}
The graphs of the function $f(x')$ and $f(f(x'))$ are shown in Fig. \ref{function_f}.

The function $f$ divides the interval $[0,\tau^2)$ into two parts.
The first one is  the interval $[0,\tau)$, window of $\Lambda_\L$, and the second
is $[\tau,\tau^2)$, window of $\Lambda_\S$. As one can see in Fig.\,\ref{function_f}
the next right neighbour of the conjugate $\tau'=-1/\tau$ of
the discontinuity point  $\tau$ is the origin $0'=0$. Next, let us
plot $f^{\lbrack 2\rbrack}(x)\equiv f(f(x))$. This function encodes the
distance to  the second next right neighbour. The interval $[0,\tau^2)$
is again split into two parts. This also means that  2-letter words
have only two possible lengths. Generally, for each $f^{\lbrack n \rbrack}$ we
will find a new point of discontinuity which splits the interval
$[0,\tau^2)$ into two parts and the $n^{\text{th}}$ next right neighbour of  its
Galois conjugate is the point 0. All points lying on the segment common to all iterates of $f$ 
up to the $n^\text{th}$ one  are the conjugates of
left-hand ends of the same $n$-letter word -- they induce a model set for this $n$-letter word.
 The following proposition easily ensues from this observation:
\begin{prop}
Left-hand ends of
all possible $n$-letter words lie as the closest $n$  left neighbours of
the origin together with the latter, namely the points $\lambda_{-n},\dots,\lambda_0$.
\end{prop}
\begin{proof}
 We know from the  above that all these points
are in different model sets and since there are $n+1$ possible words we know
 that no one is missing.
\end{proof}
If we rearrange from the largest to the smallest one all $n$-letter words
in the lexicographical order determined by
$\L>\S$,  $\Lambda^{(n)}_0>\Lambda^{(n)}_1>\cdots >
\Lambda^{(n)}_{n-1}>\Lambda^{(n)}_n$, we find that
$\Lambda^{(n)}_k=\Sigma^{[{\lambda^{(n)}_k}',\, {\lambda^{(n)}_{k+1}}')}$
for $k=0,1,\dots , n$, with
$\Lambda^{(n)}_n=\Sigma^{[{\lambda^{(n)}_n}',\tau^2)}$. The
$\lambda^{(n)}_k$'s $\in\{\lambda_{-n},\dots,\lambda_0\}$ are left-hand ends of $\Lambda^{(n)}_k$ and
are ordered as ${\lambda^{(n)}_k}' < {\lambda^{(n)}_{k +1}}' $.

Let us consider a $n$-tuple $(\lambda_k,\lambda_{k+1},\dots,\lambda_{k+n})\in\Lambda^n$. We know that $\lambda_{k+i}
-\lambda_{k+i-1}=1$ or $1/\tau$. Consequently we have
$$
\tau^2(\lambda_{k+i}-\lambda_{k+i-1}-1/\tau)=
  \begin{cases}
  \  0 & \text{if  $\lambda_{k+i}$ and $\lambda_{k+i-1}$ are ends of tile S}, \\
  \  1 & \text{if  $\lambda_{k+i}$ and $\lambda_{k+i-1}$ are ends of tile L},
  \end{cases}
$$
which implies the following property.
\begin{proper}
The number of $\L$'s in the word $[\lambda_k,\lambda_{k+n}]$ is equal to
$$\sum_{i=1}^n \tau^2(\lambda_{k+i}-\lambda_{k+i-1}-1/\tau)=\tau^2(\lambda_{k+n}-\lambda_{k}-n/\tau).$$
\end{proper}

 Let us now characterize all points of $\Sigma^{[0,1)}=\tau^2\Sigma^{[0,\tau^2)}$ in another way. From
 $$\Sigma^{[0,1)}=\left\{ l+k\tau \mid l,k\in\Z, l-\frac k\tau \in[0,1)\right\}\Longrightarrow
 \frac k\tau \leq l\leq \frac k\tau +1\Longrightarrow l=\left\lceil \frac k\tau\right\rceil,
$$
we easily recover a standard definition for chains of the Fibonacci type \cite{lest}:
\begin{proper}
$$\Sigma^{[0,\tau^2)}\equiv\frac 1{\tau^2} \Sigma^{[0,1)}\equiv\left\{ \lambda_k\ \Big| \
\lambda_k=\frac {1}{\tau^2}\left(\left\lceil \frac k\tau \right\rceil + k\tau\right),\  k\in\Z \right\}.$$
\end{proper}
It follows that the number of $\L$'s in any $n$-letter word is
$$\left(\left\lceil \frac{k+n}\tau \right\rceil+ (k+n)\tau -\left\lceil \frac k\tau\right\rceil
-k\tau  -n\tau\right)=\left\lceil\frac{k+n}\tau \right\rceil-\left\lceil \frac k\tau\right\rceil=$$
$$= \left\lceil\frac{k+n}\tau -\left\lceil \frac k\tau\right\rceil\right\rceil=\left\lceil\frac n\tau+\frac k\tau-
\left\lceil \frac k\tau\right\rceil\right\rceil=\left\lceil \frac n\tau\right\rceil \ \mbox{or} \ \left\lfloor \frac n
\tau\right\rfloor.$$ Accordingly the number of S's in any $n$-letter word is $n-\left\lceil \frac n\tau\right\rceil=
\left\lfloor \frac n {\tau^2}\right\rfloor$ or
$n-\left\lfloor \frac n \tau\right\rfloor=\left\lceil \frac n{\tau^2}\right\rceil$. In summary:
\begin{proper}\label{numberofLandS}
In any $n$-letter word there are $\left\lceil \frac n\tau\right\rceil$ $\L$'s and $\left\lfloor \frac n {\tau^2}\right\rfloor$ $\S$'s
or $\left\lfloor \frac n \tau\right\rfloor$ $\L$'s and $\left\lceil \frac n{\tau^2}\right\rceil$ $\S$'s.
\end{proper}

\section{Fibonacci B-splines and wavelets}
\subsection{B-splines scaling functions for the Fibonacci chain}
 Using   Def.\,\ref{v0} we have for $s\geq 2$,
 $${V_0^{(s)}(\Lambda)}=\!\left\{f\in C^{s-2}\!,  f\in\LR\, \big| \,
\left. f\right|_{[\lambda_n, \lambda_{n+1}]} \mbox{is polynomial
of degree} \leq s-1\!\right\}\!,$$ where we remind that $\lambda_n$
designates the $n^{\rm th}$ element of $\Lambda$ with $\lambda_0 =
0$. The Riesz basis of $V_0^{(s)}(\Lambda)$ is made up of
translates of $s+1 $ functions $\phi_{\Omega_{i}}$, where the
$\Omega_{i}$'s, $i=-s,\dots,0$, are all admissible $s$-letter
words.  Let us suppose that the support corresponding to the word
$\Omega_{i}$ is the interval $[\lambda_{i},\lambda_{i+s}]$. Then
we  construct all $s$ functions $\phi_{\Omega_{i}}\equiv \phi_i$
through the conditions below.
\begin{itemize}
\item{$\phi_i(x-\lambda_{i})=0$ if $x\notin (\lambda_{i},\lambda_{i+s})$.}
\item{$\phi_i(x-\lambda_{i})$ is  polynomial of  degree $s-1$ on the intervals
  $[\lambda_{i+k},\lambda_{i+k+1}],$ for $0 \leq k\leq s-1$.}
\item{ The $\phi_i^{(l)}(x-\lambda_{i})$'s, $\ 0\leq l\leq s-2$, are continuous in the points
$\lambda_{i+k}$ for $\ 0\leq k \leq s.$}
\item{$\displaystyle\int_{\R}\phi_i(x)\D x=\frac{\lambda_{i+s}-\lambda_{i}}{s}.$}
\end{itemize}
\subsection{Spline wavelets for the Fibonacci chain}
The construction of the wavelet basis of $W_{-1}^{(s)}(\Lambda)$ rests upon the construction of the basis
$\left\{\Psi_{n}^{(2s)},\ n\in E\right\}$ of the space $\widetilde{W}_{-1}^{(2s)}(\Lambda)$.
The first  question to answer is how many different wavelet functions
we have to build up and what are their supports.
From (\ref{setE}) and Fig.\,\ref{function_f} we  see that
$\Lambda_E=\{\lambda_n \mid n\in E\}=\left\{\lambda_n\in\Lambda \, \mid \, \lambda_{n+1}\notin\tau^2\Lambda\right\}=\Lambda_\L$.
Hence we know that all supports start with L.
 The answer about the lengths of
 supports is afforded by  Property \ref{support}. We recall that the support of the function
 $\Psi_{k}^{(2s)},\ k\in E$, is the interval $(\lambda_k,\lambda_{k+N_k})$ which includes exactly
 $2s-1$
 points from $\Lambda\backslash\tau^2 \Lambda$ and is the smallest  possible.
Let us now observe a few simple facts. We want to find all possible minimal supports containing
$2s-1$ point of $\Lambda\backslash\tau^2 \Lambda$. We know that  any $n$-letter word supports $n-1$ points of
the Fibonacci chain.  Since $\tau^2\Lambda=\Lambda_\S+1/\tau$,  the number of points of $\tau^2\Lambda$ in the
interval (word) $(\lambda_{k},\lambda_{k+n})$ is the same as the number of S's in the word
$\langle\lambda_k,\lambda_{k+n-1}\rangle$. If $\left\lceil \frac{n-1}{\tau^2}\right\rceil$ or
$\left\lfloor \frac{n-1}{\tau^2}\right\rfloor$ is the number of points of $\tau^2\Lambda$ in
the interval  $(\lambda_{k},\lambda_{k+n})$ then $\left\lfloor \frac{n-1}{\tau}\right\rfloor$ or
$\left\lceil \frac{n-1}{\tau}\right\rceil$ is the number of points of
$\Lambda\backslash\tau^2 \Lambda$ in the same interval. We now prove the following:

\begin{prop}\label{supports}
Let $p$ be the  length of the shortest word  supporting  $2s-1$
points of $\Lambda\backslash\tau^2 \Lambda$.
Then all  other minimal words
starting with $\L$  supporting $2s-1$ points of $\Lambda\backslash\tau^2 \Lambda$
have their lengths equal to $p$ or $p+1$.
\end{prop}
\begin{proof}
In  words of length $p$ there are $\left\lceil \frac{p-1}{\tau}\right\rceil=2s-1$
 or $\left\lfloor \frac{p-1}{\tau}\right\rfloor=2s-2$
points of $\Lambda\backslash\tau^2 \Lambda$. The former words are precisely those  we wish to find
while  we have to enlarge the latter with one point more. A successor  (a $(p+1)$- letter word which has the same $p$ first letters)
to words starting with L of the second type has to
support $2s-1$ points from $\Lambda\backslash\tau^2 \Lambda$ also. Indeed, if it had $2s-2$ points from $\Lambda\backslash\tau^2
\Lambda$ only, we could remove the first L (and this  removes also one point from $\Lambda\backslash\tau^2 \Lambda$). Then we
would have a $p$-letter word supporting $2s-3$ points from $\Lambda\backslash\tau^2 \Lambda$ and this is not possible.
\end{proof}

As it has been shown in the above, all words of
 length $l=\max_{k\in E}N_k$ (from Proposition \ref{supports} we also know that $l-1=\min_{k\in E}N_k$)
can be found among those having left-hand end in the interval
$[\lambda_{-l},\lambda_0].$ From the definition of $N_k$ one can infer that
there are $2s-1$ wavelets which overlap each other in the interval
$[\lambda_0,\lambda_1]$. These wavelets are certainly all different -- they have different supports.
But the first wavelet the support of which starts from $\lambda_{-l}$ or from $\lambda_{-l+1}$ (if $\lambda_{-l+1}\in\tau^2\Lambda$)
ends at $\lambda_0$ (it cannot end at $\lambda_1$ because $\lambda_{0}\in\tau^2\Lambda$ and also not at $\lambda_2$
because then the length is larger than $l$). Consequently, we have to select  this wavelet also. Hence the total number
of  different wavelets is $2s$ and we can finally assert the following:

\begin{prop}
For a given $s$ there are exactly $s+1$ B-spline scaling functions and $2s$ different wavelets. The supports of the latter  have
 length equal to  $p=\left\lceil (2s-2)\tau\right\rceil+1$ or to $p+1$.
\end{prop}

\subsection{Second-order splines  for the Fibonacci chain}
From
Fig. \ref{picture1} and also from the underlying
substitution rules, we can see that for $s=2$, there are
just 3 possible words, namely $\L\L,\, \L\S,\, \S\L$. Hence we get the following scaling
functions:
\begin{equation*}
\phi_{\L\L}(x)=\left\{ \ba{ccl} x && x\in[0,1],
\\ 2-x &&
x\in[1,2],
\ea \right.
\end{equation*}
\begin{equation*}
\phi_{\L\S}(x+\tau)=\left\{ \ba{ccl} x+\tau &&
x\in[-\tau,-1/\tau],\\ \tau^2-\tau(x+\tau) && x\in[-1/\tau,0],\ea \right.
\end{equation*}
\begin{equation*}
\phi_{\S\L}(x+1/\tau)=\left\{ \ba{ccl}
\tau(x+1/\tau) && x\in[-1/\tau,0],\\
\tau-(x+1/\tau) && x\in[0,1].\ea \right.
\end{equation*}
\begin{figure}
\begin{center}
\includegraphics[angle=0,width=10cm]{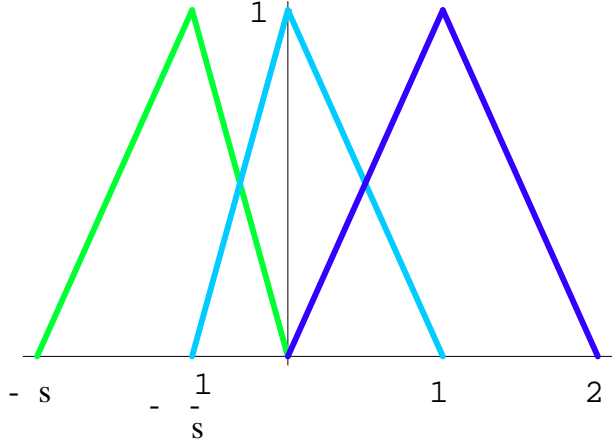}\vspace*{-.5cm}
\end{center}
\caption{Linear scaling functions $\phi_{\L\S}(x+\tau),\;
\phi_{\S\L}(x+\frac{1}{\tau})$ and $\phi_{\L\L}(x).$}
\end{figure}
The scaling equations are:\slacs{.6mm}
\begin{eqnarray*}
\phi_{\L\L}(x)&=&\frac{1}{\tau^2} \phi_{\L\L}(\tau^2
x)+\frac{2}{\tau^2}\phi_{\L\S}(\tau^2 x-1)+\phi_{\S\L}(\tau^2 x-2)
+\\
&&\frac1\tau \phi_{\L\L}(\tau^2 x-\tau^2)+\frac{1}{\tau^3}\phi_{\L\S}(\tau^2 x-\tau^2-1),\\
\phi_{\L\S}(x+\tau)&=&\frac{1}{\tau^2} \phi_{\L\L}(\tau^2
x+\tau^3)+\frac{2}{\tau^2}\phi_{\L\S}(\tau^2 x+\tau^2+
1/\tau)+ \\
&&\phi_{\S\L}(\tau^2 x+\tau+{1}/{\tau}) +\frac{1}{\tau^2} \phi_{\L\S}(\tau^2 x+\tau), \\
\phi_{\S\L}(x+{1}/{\tau})&=&\frac{1}{\tau} \phi_{\L\S}(\tau^2
x+\tau)+\phi_{\S\L}(\tau^2 x +1/\tau)+\frac{1}{\tau}
\phi_{\L\L}(\tau^2x) +\\&&\frac{1}{\tau^3} \phi_{\L\S}(\tau^2
x-1).
\end{eqnarray*}
So in this case the Riesz basis of
$V_0^{(2)}(\Lambda)$ is the set of functions
$$\left\{\phi_{\L\L}(x-\lambda_{\L\L}),\,
\phi_{\L\S}(x-\lambda_{\L\S}),\, \phi_{\S\L}(x-\lambda_{\S\L})
\right\}_{\lambda_{\L\L}\in\Lambda_{\L\L},\lambda_{\L\S}\in\Lambda_{\L\S},
\lambda_{\S\L}\in\Lambda_{\S\L}}.$$

In order to make the explicit computation of wavelets more transparent we   recall in   Table \ref{TabFib} below the points of the Fibonacci
chain $\Lambda=\{\lambda_n\}_{n\in\Z}$ around zero.

\begin{table}[!bh]
\begin{center}
\begin{tabular}{|c|c|c|c|c||c|c|c|c|c|c|}\hline
L&L&S&L&S&L&L&S&L&L&S\\
$\lambda_{-5}$&$\lambda_{-4}$&$\lambda_{-3}$&$\lambda_{-2}$&$\lambda_{-1}$&$\lambda_{0}$
&$\lambda_{1}$&$\lambda_{2}$&$\lambda_{3}$&$\lambda_{4}$&$\lambda_{5}$\\
\hline $\scsty -\tau^3$&$\scsty -\tau^2-\frac1\tau$&$\scsty
-\tau-\frac1\tau$ &$\scsty -\tau$&$\scsty -\frac1\tau$&$\scsty 0$
&$\scsty 1$&$\scsty \tau+\frac1{\tau^2}$&$\scsty \tau^2$&$\scsty
\tau^2+1$&
$\scsty \tau^3+\frac1{\tau^2}$\\
$\scsty  -1-2\tau$&$\scsty  -2\tau$&$\scsty  1-2\tau$&$\scsty
-\tau$ &$\scsty  1-\tau$&$\scsty 0 $
&$\scsty 1 $&$\scsty  2$&$\scsty  1+\tau$&$\scsty  2+\tau$&$\scsty 3+\tau $\\
yes&no&no&yes&no&yes&no&no&yes&no&no\\
\hline
\end{tabular}
\caption{ Fibonacci chain points located around the origin.} \label{TabFib}
\end{center}
\vspace{-.5cm}
\end{table}
 
 The first row tells us whether  the point is a left-hand end of the
 tile L or S, the second row is just the indexation of points, in the third
 row there  points of the Fibonacci chain are written in terms of $\tau$-expansion,
  in the fourth row they are written in the form $a+b\tau$. The last row indicates  whether the point  is element of $\tau^2 \Lambda$
 or not.

\begin{figure}[!b]
\begin{center}
\includegraphics[angle=0,width=6.5cm]{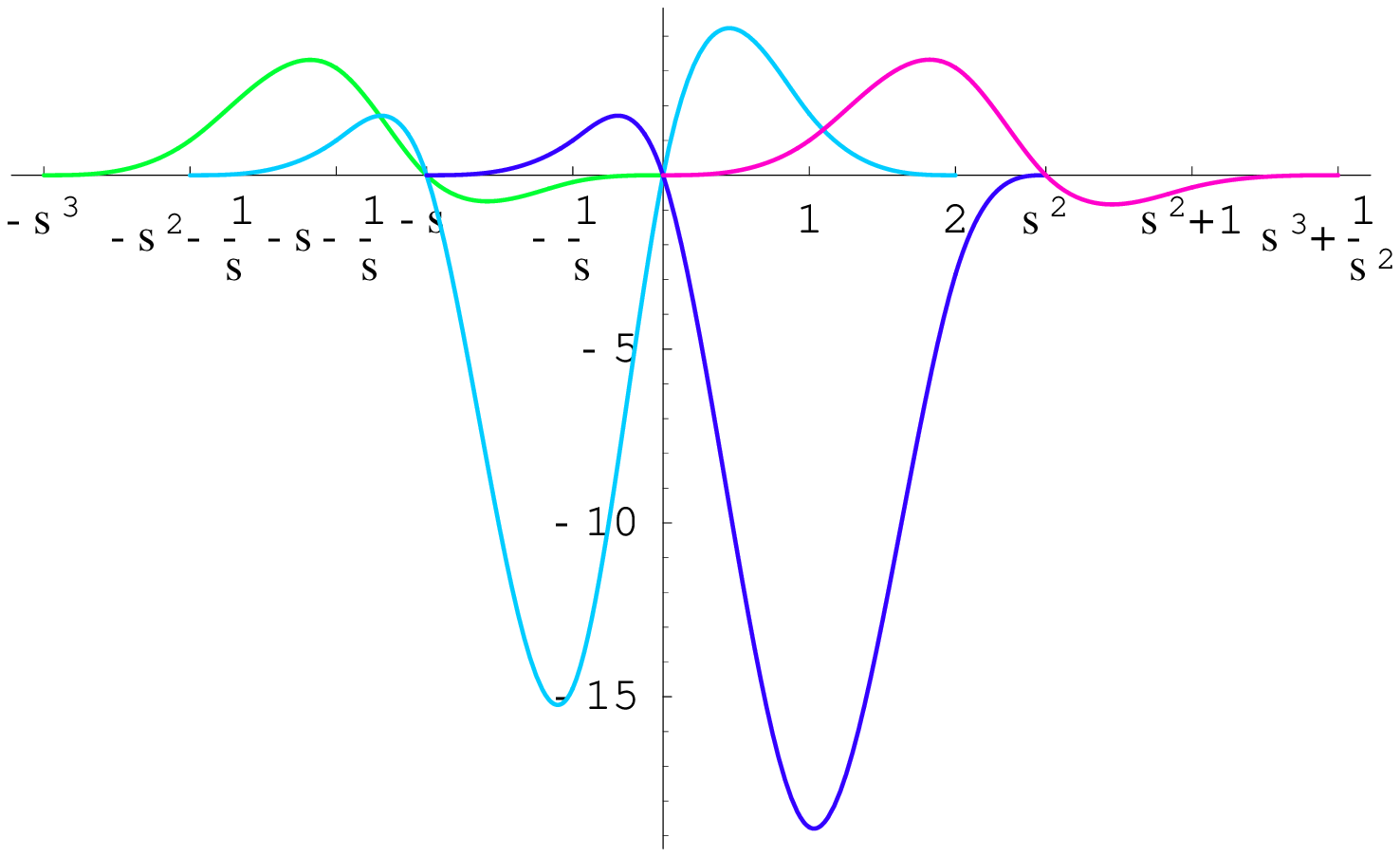}\
\includegraphics[angle=0,width=6.5cm]{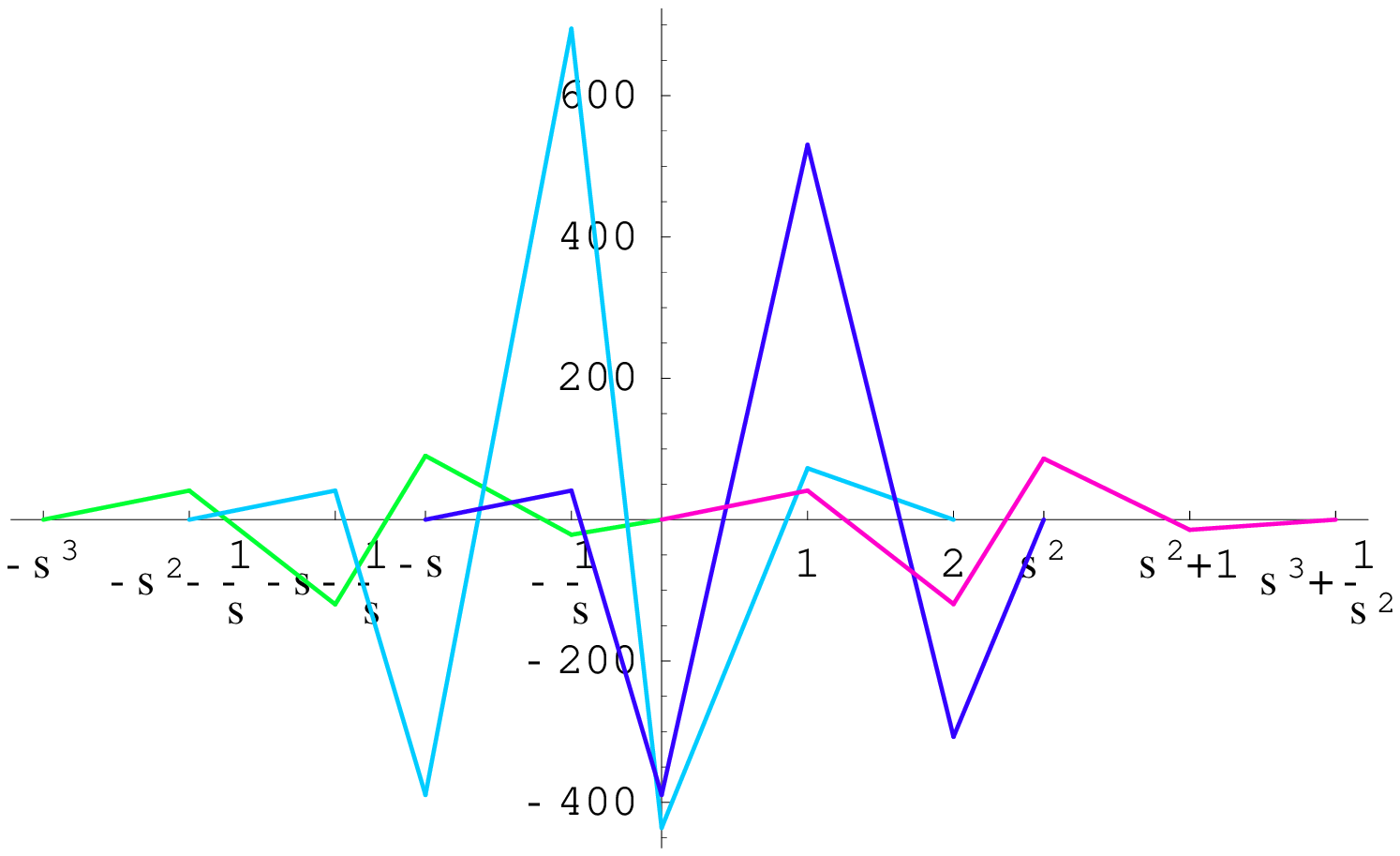}
\end{center}
\vspace{-.8cm} \caption{Four cubic functions $\Psi_n^{(4)}(x)$'s
and their second derivatives \mbox{$\zeta_{\lambda_n}(x-\lambda_n)$'s}
for $n\in\{-5,-4,-2,0\}$. } \label{fibzeta}
\end{figure}

Let us now construct the wavelet basis of $W_{0}^{(2)}(\Lambda)$
 by applying the results of Section \ref{waveletConstruction}. We
look at first for the $\Psi_n^{(2s=4)}$'s, $n\in E$, which form the
basis of $\widetilde{W}_{-1}^{(4)}(\Lambda)= \left\{f\in V_0^{(4)}
\ \big| \ f|_{\tau^2 \Lambda}=0 \right\}.$ Their respective
supports are $[\lambda_n,\lambda_{n+N_n}],$ where $n\in E\equiv \{
n\in \Z\ | \ \lambda_{n+1}\notin \tau^2 \Lambda\}$ and $N_n$ is
the smallest number for which the equality
$N_n=4+\mbox{\#}\{(\lambda_{n},\lambda_{n+N_{n}})\,\cap\,\tau^2\Lambda\}$
holds true. We recall that the equality
$\Lambda_E=\{\lambda_n\in\Lambda \mid n\in E\}=\Lambda_\L$
 determines that all supports will start by L.
The set of these supports is $ \Omega_{\psi}=\{\text{LLSLS, LSLSLL, LSLLS,
LLSLL}\}$. This means that
 we have four different shapes of $\Psi_n^{(4)}$'s, one per  support.
 As  second derivatives of functions $\Psi_n^{(4)}(x),\ n\in E$ we obtain
 four different functions $\zeta_{\lambda_n}(x)$,
 precisely $\zeta_{\lambda_n}(x-\lambda_n)=\frac{\D^2\phantom{x}}{\D x^2}\Psi_n^{(4)}(x),\ n\in E$ (see Fig.~\ref{fibzeta}).
 Due to the equality  $\zeta_{\lambda_n}(x)=\zeta_{\lambda_k}(x)$ if $\lambda_n,\ \lambda_k\in\Lambda_\mu$
  where $\mu\in  \Omega_{\psi}$, we just denote the latter by $\zeta_\mu(x)$.

\begin{table}[!t]
\begin{center}
\begin{tabular}{|c|l|c|c|c|c|}\hline\rule{0pt}{.5cm}
&&\multicolumn{2}{|c|}{$\zeta_{\rm LLSLS}(x-\lambda_{-5})$}&
\multicolumn{2}{|c|}{$\zeta_{\rm LSLSLL}(x-\lambda_{-4})$}
  \tabularnewline[1mm]
 \cline{3-6}\rule{0pt}{.5cm}
 tile&interval&$k_i$&$q_i$&$k_i$&$q_i$ \tabularnewline[1mm]\hline\rule{0pt}{.5cm}
 L&$[\lambda_{-5},\lambda_{-4})$&$\scsty 6$&$\scsty 0$&&\tabularnewline[1mm]
L&$[\lambda_{-4},\lambda_{-3})$&$-\frac{6(1+26\tau)}{11}$&$\scsty
6$&
 $\scsty 6$&$\scsty 0$\tabularnewline[1mm]
S&$[\lambda_{-3},\lambda_{-2})$&
$\frac{12(18+17\tau)}{11}$&$\frac{12(5-13\tau)}{11}$
 &$\scsty -24(1+2\tau)$&$\scsty 6$\tabularnewline[1mm]
L&$[\lambda_{-2},\lambda_{-1})$&
$-\frac{24(1+4\tau)}{11}$&$\frac{12(4+5\tau)}{11}$
 &$\frac{6(91+123\tau)}{11} $&$\scsty -6(3+4\tau)$\tabularnewline[1mm]
 S&$[\lambda_{-1},\lambda_{0})$&
$\frac{12(3+\tau)}{11}$&$\frac{12(2-3\tau)}{11}$
 &$-\frac{6(127+224\tau)}{11} $&$\frac{6(58+79\tau)}{11} $\tabularnewline[1mm]
L&$[\lambda_{0},\lambda_{1})$& &
 &$\frac{21(13+16\tau)}{11} $&$-\frac{18(13+16\tau)}{11} $\tabularnewline[1mm]
L&$[\lambda_{1},\lambda_{2})$& &
 &$-\frac{3(13+16\tau)}{11} $&$\frac{3(13+16\tau)}{11} $\tabularnewline[2mm]
 \hline\hline
  \multicolumn{2}{|c|}{norm} & \multicolumn{2}{|c|}{$\scsty\frac{2}{11}\,\sqrt{6(270+431\tau)}\doteq 13.8519$}
  & \multicolumn{2}{|c|}{\rule{0pt}{.5cm}$\scsty\frac{2}{11}\,\sqrt{3(16754+27145\tau)}\doteq 77.572$}  \tabularnewline[2mm]
 \hline
\end{tabular}
\caption{Analytical expression of $\zeta_{\rm
LLSLS}(x-\lambda_{-5})$ and $\zeta_{\rm LSLSLL}(x-\lambda_{-4})$.
} \label{TabWav1} \vspace{-.5cm}
\end{center}
\end{table}
From  Table \ref{TabWav1} and \ref{TabWav2} we  deduce
the analytical expressions of functions $\zeta_\mu$, $\mu\in  \Omega_{\psi}$
translated to their basic  admissible  intervals. On each
subinterval $[\lambda_i,\lambda_{i+1})$ the function
$\zeta_\mu(x-\lambda_\mu)$  is described by a linear function
$k_i(x-\lambda_i)+q_i$. The values of $k_i$'s and $q_i$'s are given in
these tables.

\begin{remark}
It should be noticed from all elements of the extension field $\Q(\tau)$ shown in the tables that imposing  $\Psi_n^{(4)}(\lambda_{n+1}) = 1$  is not an optimal choice. The simplest expressions are found with the choice
$\Psi_n^{(4)}(\lambda_{n+1}) = 11/3$. The appearing of number $11$ is due to an interesting algebraic feature. In order to determine the coefficients of the linear $\zeta$'s we have to solve a system of equations with coefficients in the field $\Q(\tau)$, and so all solutions are of the form $\frac{v}{w} \equiv \frac{a + b \tau}{c + d \tau} = \frac{(a + b \tau)(c - d /\tau)}{c^2 - d^2 +cd} \in \Q(\tau)$. Now, it happens that  the  ``algebraic squared norms''
$ww' = c^2 - d^2 + cd$ of denominators are equal to $\pm 1$ (\textit{i.e.} $w$ is unit in $\Z \lbrack \tau \rbrack$) or to  $\pm 11$. More generally, properties and shapes of wavelets are independent of the fixing $\Psi_n^{(4)}(\lambda_{n+1}) = u$. With the generic latter choice, all values in the tables  \ref{TabWav1} and \ref{TabWav2} should be multiplied by~$u$.
\end{remark}

\begin{table}[!t]
\begin{center}
\begin{tabular}{|c|l|c|c|c|c|}\hline\rule{0pt}{.5cm}
&&\multicolumn{2}{|c|}{$\zeta_{\rm LSLLS}(x-\lambda_{-2})$}&
\multicolumn{2}{|c|}{$\zeta_{\rm LLSLL}(x-\lambda_{0})$}
  \tabularnewline[1mm]
 \cline{3-6}\rule{0pt}{.5cm}
 tile&interval&$k_i$&$q_i$&$k_i$&$q_i$ \tabularnewline[1mm]\hline\rule{0pt}{.5cm}
 L&$[\lambda_{-2},\lambda_{-1})$&$\scsty 6$&$\scsty 0$&&\tabularnewline[1mm]
S&$[\lambda_{-1},\lambda_{0})$&$\scsty -24(1+2\tau)$&$\scsty 6$&
 &\tabularnewline[1mm]
L&$[\lambda_{0},\lambda_{1})$& $\scsty 3(14+19\tau)$&$\scsty
-6(3+4\tau)$
 &$\scsty 6$ &$\scsty 0$\tabularnewline[1mm]
L&$[\lambda_{1},\lambda_{2})$& $\scsty -3(10+19\tau) $&$\scsty
3(8+11\tau)$
 &$-\frac{6(-25+42\tau)}{11} $&$\scsty 6$\tabularnewline[1mm]
 S&$[\lambda_{2},\lambda_{3})$&
$\scsty 6(4+5\tau)$&$\scsty -6(1+4\tau)$
 &$\frac{36(10+3\tau)}{11} $&$-\frac{36(-6+7\tau)}{11}$\tabularnewline[1mm]
L&$[\lambda_{3},\lambda_{4})$& &
 &$-\frac{42(-1+3\tau)}{11} $&$\frac{36(-1+3\tau)}{11}$\tabularnewline[1mm]
L&$[\lambda_{4},\lambda_{5})$& &
 &$\frac{6(-1+3\tau)}{11} $&$-\frac{6(-1+3\tau)}{11} $\tabularnewline[2mm]
 \hline\hline
 \multicolumn{2}{|c|}{norm}  & \multicolumn{2}{|c|}{$\scsty{\sqrt{6(191+309\tau)}}\doteq 64.3882$}
  & \multicolumn{2}{|c|}{\rule{0pt}{.5cm}$\scsty\frac{2}{11}\,\sqrt{6(281+411\tau)}\doteq 13.6981$} \tabularnewline[2mm]  
  \hline
\end{tabular}
\caption{Analytical expression for $\zeta_{\rm
LSLLS}(x-\lambda_{-2})$ and $\zeta_{\rm LLSLL}(x-\lambda_{0})$.}
\label{TabWav2} \vspace{-.5cm}
\end{center}
\end{table}

\begin{figure}[!bh]
\begin{center}
\includegraphics[angle=0,width=10cm]{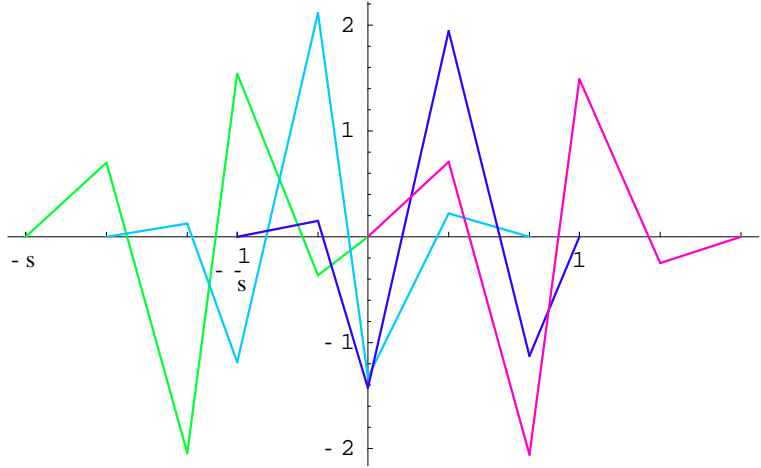}
\end{center}
\vspace{-.8cm} \caption{Basic admissible translations of four
linear wavelets $\psi_\mu,\ \mu\in  \Omega_{\psi}$.} \label{fibwave}
\end{figure}

By
 dilation and normalization of the $\zeta_\mu(x)$'s we eventually obtain four mother wavelets $\psi_\mu(x)=
 \frac{\zeta_{\mu}(\tau^2 x)}{||\zeta_{\mu}(\tau^2 x)||}$,
$\mu\in  \Omega_{\psi}$, see Fig.~\ref{fibwave} for their basic admissible
translations. Then the set of functions
$\{\psi_\mu(x-\kappa_\mu)\}_{\mu\in  \Omega_{\psi}, \kappa_\mu\in
\tau^{-2}\Lambda_\mu}$ forms a Riesz basis of $W_0^{(2)}(\Lambda)$
and consequently the set of functions
$$
\left\{ \tau^j \psi_\mu (\tau
^{2j}x-\kappa_\mu)\right\}_{j\in\Z,\,\mu\in \Omega_{\psi},\,\kappa_\mu\in\tau^{-2}\Lambda_\mu}
$$
forms a Riesz basis of $\LR$.

\subsection{Construction of  scaling equations for wavelets}
Let us here describe  the algorithm for obtaining scaling equations  in the case of piecewise linear wavelets. 
For convenience we first set up the equations for the
functions $\zeta_\mu,\ \mu\in \Omega_{\psi}$. The corresponding 
 equations for wavelets $\psi_\mu,\ \mu\in \Omega_{\psi}$ are trivially derived from them. Let us 
 denote by $\Omega_{\phi} $ the set of words supporting the scaling functions (in the present case $\Omega_{\phi} =  \{\L\L,\L\S,\S\L\}$). Due to
$W_{-1}^{(2)}\subset V_0^{(2)}$ we can  write
$$
\zeta_\mu(x-\lambda_\mu)=\sum_{\nu \in \Omega_{\phi} ,l\in\Lambda_\nu}g_{\nu,l}^\mu
\phi_\nu( x-l).
$$
 The translate to the point
$\lambda_i\in\Lambda_\nu$ of a scaling function $\phi_\nu$ for all
$\nu\in \Omega_{\phi} $ reads as:
$$
\phi_{\lambda_i}(x-\lambda_i)=\left\{%
\begin{array}{ll}\disty
    \frac{x-\lambda_i}{\lambda_{i+1}-\lambda_i}, & \hbox{for }  x\in [\lambda_i,\lambda_{i+1}],\\[4mm]
    \disty\frac{\lambda_{i+2}-\lambda_i-(x-\lambda_i)}{\lambda_{i+2}-\lambda_{i+1}}=
    \frac{\lambda_{i+2}-x}{\lambda_{i+2}-\lambda_{i+1}},\ & \hbox{for }
      x\in [\lambda_{i+1},\lambda_{i+2}].
\end{array}%
\right.
$$

 We look for $g_i$'s in the
scaling equation \be
\zeta_{\lambda_i}(x-\lambda_i)=\sum_{j=i}^{i+N_i-2}g_j
\phi_{\lambda_j}(x-\lambda_j). \ee
 $\bullet \ \ \bmth{{[\lambda_i,\lambda_{i+1}]}:}$ \\
 Because of support condition and  continuity   at the point $\lambda_i$ we can write
 $$k_i(x-\lambda_i)+q_i \big|_{x=\lambda_i}=0,$$
and so the coefficient $q_i$ has to be 0.  Hence we have on this interval the
equation
$$
k_i(x-\lambda_i)=g_i \frac{x-\lambda_i}{\lambda_{i+1}-\lambda_i}
\Longrightarrow g_i=k_i(\lambda_{i+1}-\lambda_i).
$$
The function is continuous at $\lambda_{i+1}$ so we have
$$k_i(x-\lambda_i)\big|_{x=\lambda_{i+1}}=k_{i+1}(x-\lambda_{i+1})+q_{i+1}\big|_{x=\lambda_{i+1}}, $$
thus \be\label{q_i+1} q_{i+1}=k_i(\lambda_{i+1}-\lambda_i)=g_i.
\ee
$\bullet \ \ \bmth{{[\lambda_{i+1},\lambda_{i+2}]}:}$\\
On the interval ${[\lambda_{i+1},\lambda_{i+2}]}$ we have the equation
$$
k_{i+1}(x-\lambda_{i+1})+q_{i+1}=g_{i}
\frac{\lambda_{i+2}-x}{\lambda_{i+2}-\lambda_{i+1}}+
g_{i+1}\frac{x-\lambda_{i+1}}{\lambda_{i+2}-\lambda_{i+1}}.
$$
Using the equation (\ref{q_i+1}) we obtain $
g_{i+1}=k_{i+1}(\lambda_{i+2}-\lambda_{i+1})+q_{i+1}. $ From the
continuity at the point $\lambda_{i+2}$ we also have
$$k_{i+1}(x-\lambda_{i+1})+q_{i+1}\big|_{x=\lambda_{i+2}}=k_{i+2}(x-\lambda_{i+2})+q_{i+2}\big|_{x=\lambda_{i+2}},$$
thus $q_{i+2}=g_{i+1}.$\\[2mm]
$\bullet \ \ \bmth{{[\lambda_{i+l},\lambda_{i+l+1}]}:}$\\
By induction we have generally for $l,\ 0\leq l
\leq N_i-2$ \be
g_{i+l}=k_{i+l}(\lambda_{i+l+1}-\lambda_{i+l})+q_{i+l}=q_{i+l+1}.
\ee This method is simple and very general. Actually it can be carried out for any
self-similar locally finite Delone set in the case of piecewise linear wavelets.

\begin{table}
\sltcs{1mm}
\begin{center}
\begin{tabular}{|c|c|c|c|c|}
\hline &\rule{0pt}{.5cm}  $\scsty\psi_{\rm LLSLS}(x-\kappa_{-5})$& $\scsty\psi_{\rm
LLSLS}(x-\kappa_{-4})$&$\scsty\psi_{\rm LSLSLL}(x-\kappa_{-2})$ &
$\scsty\psi_{\rm LLSLL}(x)$\tabularnewline[1mm]
\hline\rule{0pt}{.5cm}
$\phi_{\L\L}(\tau^2 x -\lambda_{-5})$& $ 6$& & &\tabularnewline[1mm]
\hline\rule{0pt}{.5cm}
$\phi_{\L\S}(\tau^2 x -\lambda_{-4})$& $  \frac{12}{11}(5-13\tau)$& $ 6$& &\tabularnewline[1mm]
\hline\rule{0pt}{.5cm}
$\phi_{\S\L}(\tau^2 x -\lambda_{-3})$& $  \frac{12}{11}(4+5\tau)$& $  -6(3+4\tau)$& $   $&\tabularnewline[1mm]
\hline\rule{0pt}{.5cm}
$\phi_{\L\S}(\tau^2 x -\lambda_{-2})$& $  \frac{12}{11}(2-3\tau)$& $  \frac{6}{11}(58+79\tau)$& $  6$&\tabularnewline[1mm]
\hline\rule{0pt}{.5cm}
$\phi_{\S\L}(\tau^2 x -\lambda_{-1})$&  & $  -\frac{18}{11}(13+16\tau)$& $  -6(3+4\tau)$&\tabularnewline[1mm]
\hline\rule{0pt}{.5cm}
$\phi_{\L\L}(\tau^2 x -\lambda_{0})$&  & $\frac{3}{11}(13+16\tau)$& $  3(49+79\tau)$&$6$\tabularnewline[1mm]
\hline\rule{0pt}{.5cm}
$\phi_{\L\S}(\tau^2 x -\lambda_{1})$&  &  & $  -6(1+4\tau)$&$\frac{36}{11}(6-7\tau)$\tabularnewline[1mm]
\hline\rule{0pt}{.5cm}
$\phi_{\S\L}(\tau^2 x -\lambda_{2})$&  &  &  &$-\frac{36}{11}(1-3\tau)$\tabularnewline[1mm]
\hline\rule{0pt}{.5cm}
$\phi_{\L\L}(\tau^2 x -\lambda_{3})$&  &  &  &$\frac{6}{11}(1-3\tau)$\tabularnewline[1mm]
\hline
\end{tabular}
\caption{Table of coefficients for scaling equations of
$\psi_\mu,\ \mu\in  \Omega_{\psi}.$} \label{tab-scal-eq-wav-fib}
\end{center}
\vspace{-.8cm}
\end{table}
Finally the scaling equations for basic admissible translations of
functions $\psi_\mu,$ $ \mu\in  \Omega_{\psi}$
 have the form
$$\psi_\mu(x-\kappa_\mu)=\sum_{\nu\in  \Omega_{\phi},\,l\in\Lambda_\nu}g_{\nu,l}^\mu
\frac{\phi_\nu(\tau^2 x-l)}{||\zeta_\mu(\tau^2 x)||}, \  \kappa_\mu \in \theta^{-1}\LA_{\mu}.$$
Coefficients $g_{\nu,l}^\mu$ are recalled in Table
\ref{tab-scal-eq-wav-fib}. The norms in denominators are:
\begin{align*}
||\zeta_{\L\L\S\L\S}(\tau^2 x)||&=\frac{2}{11}\sqrt{6(109+161\tau)}\doteq 8.561,\\
||\zeta_{\L\S\L\S\L\L}(\tau^2 x)||&=\frac{2}{11}\sqrt{3(6363+10391\tau)}\doteq 47.942,\\
||\zeta_{\L\S\L\L\S}(\tau^2 x)||&=\sqrt{6(73+118\tau)}\doteq 39.794,\\
||\zeta_{\L\L\S\L\S}(\tau^2
x)||&=\frac{2}{11}\sqrt{6(151+130\tau)}\doteq 8.466.
\end{align*}

\section{Wavelets for stone-inflation tilings}
\label{multidim}
\begin{figure}
\includegraphics[angle=0,width=13cm,clip=]{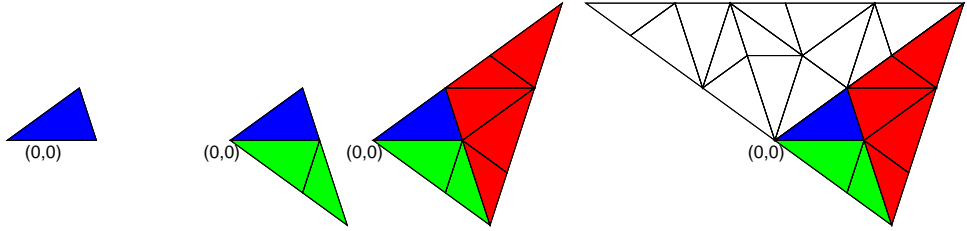}
\caption{First three steps of stone-inflation for a Penrose tiling.} \label{penrose}
\end{figure}

The construction of spline wavelets for self-similar Delaunay point sets in $\R^d$ has also been considered by Bernuau in \cite{bern1}.
 Here we shall briefly describe similar  constructions for  a tiling
$\T$ of $\R^d$ which has the  \emph{stone-inflation} symmetry $\sigma\T\subset\T$ where
 $\sigma=(b,\theta R)\in \R^d\rtimes \left(\R^*_+\times SO(d) \right)$, $\theta>1$,  is an affine transformation
 of $\R^d$. Related
constructions were also described in \cite{gazkrej,gazkram}.
Let us first recall  what we understand by tiling. A tiling of
$\R^d$ is a covering with non-overlapping pieces all congruent (in
the Euclidean group action sense) to tiles belonging to a
predetermined finite set of {\em prototiles} or {\em tile alphabet},
$\mathcal{A}=\lbrace T_1, \dots, T_n \rbrace$:
\begin{equation}
\R^d = \bigcup_{i\in \N} P_i, \ P_i \stackrel{\text{cong}}{=} T_l,
\  T_l \in\mathcal{A}, \ P_i \cap P_j \subset \R^{d'}, \ i\neq j,
\, d' < d.
\end{equation}
In this definition, a tile is supposed to be compact, equal to the closure of its interior, and homeomorphic to a topological ball. Moreover, in the tiling any two tiles have pairwise disjoint interiors (all tiles of the tiling  \emph{pack} $\R^d$).
Many tilings $\mathcal{T}$ of $\R^d$ have the so-called stone-inflation
symmetry: all tiles of $\mathcal{T}$, when rotated and scaled by a
given $\theta R  > 1$, $\theta >1$, $R \in SO(d)$, and translated
by a given $b \in \R^d$,  can be packed face-to-face from the
original ones. An example is provided by a tiling in the Penrose class as shown in Fig.~\ref{penrose}.  Another example, also relevant to quasicrystalline studies is the three-dimensional Danzer tiling \cite{ankra}.

More precisely, suppose we are given a tiling
$\mathcal{ T}$ of $\R^d$ built from a finite set of prototiles $\{
T_1, \dots, T_n\}$ present in the tiling,
\begin{equation}\label{new1}
\mathcal{ T} = \bigcup_{i\in \N} P_i=\bigcup_{l=1}^{n}
\bigcup_{\gamma_l \in \Gamma_l} \gamma_l \cdot T_l,
\end{equation}
where $\gamma_l = (b_l, R_l)$ is a translation-rotation in the
Euclidean group $\R^d \rtimes SO(d)$ and $\Gamma_l \subset \R^d
\rtimes SO(d)$ is made of all those transformations $\gamma_l =
(b_l, R_l)$ (including the identity) which bring the prototile $T_l$ to one of its
congruent companions appearing in the tiling. Here, we keep the
freedom to restrict the $\Gamma_l$'s to pure translation sets, at
the price of enlarging the number of prototiles. The
stone-inflation  symmetry  based on the affine-linear inflation
$\sigma = (b, \theta R)$ then means that for each $l$ and each
$\gamma_l \in \Gamma_l$, the following finite patch
\begin{equation}
\sigma P_i=\sigma \gamma_l \cdot T_l = \bigcup_{m=1}^n
\bigcup_{\sigma_{lm} \in \Gamma_m}\sigma_{lm} \cdot T_m,
\end{equation}
is present in the tiling. Hence, it becomes possible to deal with
an infinite sequence of inflated-deflated versions of the tiling
$\T$
 \be
 \label{tiling}
  \cdots \T_{j-1}\subset \sigma^{-j} \T \stackrel{\text{def}}{=}
\bigcup_{i\in \N} \sigma^{-j} \cdot P_i
\stackrel{\text{def}}{=} \T_j\subset \T_{j+1} \cdots . 
\ee
Here, the inclusion relation between two tilings should be understood as: 
$\T \subset \T'$ if any tile of $\T$ is patch of tiles of $\T'$. Let $\Lambda (\T)$ denote the set of vertices of the tiling $\T$. It is clearly Delaunay and we have the counterpart of (\ref{tiling}):
\begin{equation}
\label{tilingvert}
 \cdots \LA(\T_{j-1}) \subset \LA(\T_{j})\subset\LA(\T_{j+1}) \cdots . 
\end{equation}
Furthermore, we have the denseness property:
\begin{equation}
\label{ }
\R^d=\overline{\bigcup_j \LA(\T_j}).
\end{equation}

\subsection{Haar wavelet basis}
In order to prepare the discussion about the feasibility of finding  spline wavelets adapted to tilings of the above type, let us  first adapt to our context the  construction of  an
orthonormal Haar wavelet basis of $L^2(\R^d)$ proposed  in \cite{bern1}.  We first define a sequence of spaces $(V_j(\T))_{j\in\Z}$
such that each space  $V_j(\T)$ is the closed subspace of
$L^2(\R^d)$ of functions which are constant on all tiles
$\sigma^{-j}P_i$, $i\in\N$. Then we have,
\begin{prop}\label{n-multiresolution}
The sequence $(V_j(\T))_{j\in\Z}$ is a {\em
$\sigma$-multiresolution analysis of $L^2(\R^d)$} in the following sense:
\begin{itemize}\itemsep 0mm
    \item [(i)] for all $j\in\Z$, $V_{j-1}(\T)\subset V_j(\T),$
    \item [(ii)] $\bigcup_{j\in\Z} V_j(\T)$ is dense in $L^2(\R^d)$,
    \item [(iii)] $\bigcap_{j\in\Z}V_j(\T)=\{0\},$
    \item [(iv)] for all $j\in\Z, \x\in\R^d$, $f(\x)\in V_j(\T)\Longleftrightarrow f(\sigma ^{-j}\x)\in V_0(\T),$
    \item [(v)] there exists a finite number $n$ of  scaling functions, $\phi_1(\x),\phi_2(\x),\dots,\phi_n(\x)$ such that
    all their admissible linear-affine transformations $ \{ \phi_l (\gamma_l^{-1}\cdot \x)\}_{1\leq l
\leq n, \, \gamma_l \in \Gamma_l}$  form an orthonormal  basis in $V_0(\T)$.
\end{itemize}
\end{prop}
\begin{proof}\indent
\begin{itemize}\itemsep 0mm
\item [(i)]  This inclusion  results from the stone-inflation property of tiling $\T$, $\sigma \T\subset \T$.
\item [(ii)] This is true through the fact that
 every continuous function $f$ with  compact support on $\R^d$ can be written as
 uniform limit of the sequence $(f_j)_{j\geq 0}$,
$$f_j(\x)=\sum_{m\in\N}f(\x_{j,m}){\bf 1}_{\sigma^{-j}P_m}(\x),$$ where
$\x_{j,m}\in \sigma^{-j}P_m$.
 \item [(iii)] By  construction it is clear that only the function $f(\x)=0$
 is included in all spaces $V_j(\T)$.
\item [(iv)] Let us choose $f(\x)\in V_j(\T)$. Then we have
$$
f(\x)=\sum_{m\in\N}c_m {\bf 1}_{\sigma^{-j}P_m}(\x), \
\text{with}\ \sum_{m\in \N}|c_m|^2<\infty,
$$
and this is equivalent to
$$
f(\sigma^{-j}\x)=\sum_{m\in\N}c_m {\bf 1}_{P_m}(\x).
$$
Thus $f(\sigma^{-j}\x)\in V_0(\T)$.
\item[(v)]
We have $n$ scaling functions
$\phi_1(\x),\phi_2(\x),\dots,\phi_n(\x) $ which are the normalized
characteristic functions  of the corresponding tiles $T_l,\
l=1,\dots,n$ divided by the square root of its volume, \emph{e.g.} for a tile
$X\in\mathcal{A}$
$$\phi_X(\x)={\bf 1}_X(\x)/\sqrt{|X|} $$
where $|X|$ means the volume of the tile $X$.  Due to 
normalization we see that all admissible linear-affine
transformations of $n$ functions $\{\phi_l\}_{l=1}^n$ form an
orthonormal basis of $V_0(\T)$.
\end{itemize}
\end{proof}

We now proceed to the construction of the Haar wavelets.  For each prototile $T_l$,
$1\leq l\leq n$ there are finitely many tiles $P_{l_j}$, such that:
$$
\sigma T_l=P_{l_1} \cup \cdots\cup P_{l_{k_l}} \Longleftrightarrow
T_l=\sigma^{-1}P_{l_1} \cup \cdots\cup \sigma^{-1}P_{l_{k_l}},
$$
and the set $\left\{ P_{l_1},\dots,P_{l_{k_l}} \right\}$ forms a patch present in the tiling $\T$.
 Denote by $V_{0,P_i}(\T)$ the subspace of $V_0(\T)$ of functions
which are zero (almost everywhere) outside  the tile $P_i$. Accordingly we define
 $V_{j,P_i}(\T)$ as the subspace of $V_j(\T)$
of functions equal to zero  (almost everywhere) outside  the tile $P_i$.
Therefore we obtain the  orthogonal decompositions:
\begin{align}
V_0(\T)&=\bigoplus_{i\in\N}\,\Bot V_{0,P_i}(\T),\\
V_j(\T)&=\bigoplus_{i\in\N}\,\Bot V_{j,P_i}(\T),
\end{align}
and the inclusions
\begin{equation}
 V_{0,P_i}(\T)\subset V_{1,P_i}(\T)\subset\cdots
\subset V_{j,P_i}(\T)\cdots.
\end{equation}
The wavelet space $W_0(\T)$ is the orthogonal complement of
$V_0(\T)$ in $V_1(\T)$,
\begin{equation}
V_1(\T)=V_0(\T)\oplus_{\!\bot} W_0(\T).
\end{equation}
More generally $$V_{j+1}(\T)=V_j(\T)\oplus_{\!\bot} W_j(\T).$$ We
 also define  the orthogonal complement of $V_{j,P_i}(\T)$ in
$V_{j+1,P_i}(\T)$ as
\begin{equation}
V_{j+1,P_i}(\T)=V_{j,P_i}(\T)\oplus_{\!\bot} W_{j,P_i}(\T).
\end{equation}
It is clear that
$$
W_j(\T)=\bigoplus_{i\in\Z}\,\Bot W_{j,P_i}(\T).
$$
Finally we get the orthogonal decomposition of the whole Hilbert space:
$$
L^2(\R^d)=\bigoplus_{j\in\Z}\,\Bot W_j(\T).
$$
Thus the construction of $W_0(\T)$ is equivalent to the
construction of all $W_{0,P_i}(\T)$'s. Since any tile $P_i$ can be written
as
$$
P_i=\gamma_lT_{l}=\gamma_l(\sigma^{-1}P_{l_{1}} \cup \cdots\cup
\sigma^{-1}P_{l_{k_{l}}}),
$$
 it is sufficient to find wavelets for prototiles  $T_l,\
l=1,\dots,n$, and the whole basis of $W_0(\T)$ will be formed by all
admissible linear-affine transformations of these ``protowavelets''. There results
the following proposition.
\begin{prop}\label{n-Haar-wav}
For every prototile $T_l,\ l=1,\dots ,n$, given also by
$$
T_l=\sigma^{-1}P_{l_1} \cup \cdots\cup \sigma^{-1}P_{l_{k_l}},
$$
we have $k_l-1$ orthonormal wavelets
$$
\psi_{1,l}(\x),\dots,\psi_{k_{l}-1, l}(\x).
$$
\end{prop}
\begin{proof}
Let us denote by $V_{0,T_l}(\T)$ the space of functions constant on tile
$T_l$ and equal to zero otherwise, \emph{i.e.} those functions proportional to $
{\bf 1}_{T_l}(\x)$.  We then denote by
$V_{1,T_l}(\T)$ the space of functions constant on tiles
$\sigma^{-1} P_{l_1},\dots,\sigma^{-1}P_{l_{k_l}}$ and otherwise
equal to zero. The space of wavelets corresponding to the tile
$T_l$ is found as the orthogonal complement of $V_{0,T_l}(\T)$ in
$V_{1,T_l}(\T)$, i.e.
$$
V_{1,T_l}(\T)=V_{0,T_l}(\T)\oplus_{\!\bot} W_{0,T_l}(\T).
$$
One possible construction of  an orthonormal basis of $W_{0,T_l}(\T)$ is as follows:
we start from the set of functions $\{{\bf 1}_{T_l}(\x),{\bf
1}_{\sigma^{-1} P_{l_1}}(\x), \dots,{\bf 1}_{\sigma^{-1}
P_{l_{k_l-1}}}(\x)\}$ which form a basis of $V_{1,T_l}(\T)$.
Because of the presence of the function ${\bf 1}_{T_l}(\x)$ we have put aside (for instance) the last function ${\bf
1}_{\sigma^{-1} P_{l_{k_l}}}(\x)$. We then  proceed to
 Gram-Schmidt orthogonalization and $L^2$-normalization. Thus we get an orthonormal basis of $V_{1,T_l}(\T)$ and
after removing the single basis element in $V_{0,T_l}(\T)$ we are left with
$k_l-1$ functions which form an orthonormal basis of
$W_{0,T_l}(\T)$.
\end{proof}

The orthonormal Haar basis of $L^2(\R^d)$ adapted to the tiling $\T$ is finally given by:
\begin{equation*}
\bigcup_{j \in \Z}\bigcup_{1 \leq l \leq n}\bigcup_{\gamma_{l} \in \Gamma_{l}} 
\left\{ \theta^{j\frac{d}{2}} \psi_{l_1,l}\left( \gamma_{l}^{-1} \sigma^j \cdot \x \right), \dots ,  \theta^{j\frac{d}{2}} \psi_{l_{k_l - 1},l}\left( \gamma_{l}^{-1} \sigma^j \cdot \x \right)\right\}.
\end{equation*}

\subsection{Beyond Haar: spline wavelet basis}
Let $\T$ be a stone inflation tiling of $\R^d$. Suppose prototiles are polytopes and  the set $\Lambda(\T)$ of vertices is of 
 finite local complexity.  How to build piecewise linear, compactly supported, spline scaling functions and related wavelets living on tiles of $\T$?  More precisely, one could think about going through the following steps. 
\begin{itemize}
  \item How many tiles are needed for building the minimal patches supporting scaling functions? 
  \item How many different patches of such type exist in the tiling?
  \item How to characterize such spline functions from a functional analysis point of view?
   \item Suppose the previous questions answered. How to build the corresponding multiresolution?
   \item Suppose the previous question be solved and related scaling functions be determined. How to find the corresponding wavelets?
   \item Are these wavelets compactly supported? 
   \item If yes, on which patches in the tiling?
   \item How to extend this material to smoother splines?
   \end{itemize}
   
   It is not the aim of the present paper to achieve this ambitious program. Let us just sketch which procedure  could be followed in the simplest two-dimensional case.
   In order to build ``pyramidal''  functions (the two-dimensional analogue of the one dimensional spline ``hat''
  functions) which are to play the role of  scaling functions,  we have to accomplish
   a triangulation of the tiling $\T$ (like we already have for Penrose tilings). By  triangulation
   we mean that in each prototile some of the vertices could be pairwise connected with a new segment (in order to divide the prototile into
   triangles) under the condition  not to create
   new nodes (the new edges should not cross). All tiles congruent to the divided prototile will be
   triangulated in the same manner. Then
to every point  $\lambda\in\Lambda(\T)$ there is an associated  pyramid. The support of this pyramid is delimited
   by the  neighbours of $\lambda$. (A neighbour is a point connected with $\lambda$ by an edge). The vertex of the pyramid is obviously located right above
   $\lambda$. Because of the finite local complexity  we know that there  exists
   a finite number of different pyramids. These pyramidal functions   are uniquely
   determined once  a certain normalization condition is fulfilled.
   These pyramids and all their admissible linear-affine transformation
then define  the space $V_0(\T)$. The multiresolution analysis follows.

\begin{figure}[!th]
\begin{center}
\includegraphics[angle=0,width=9cm]{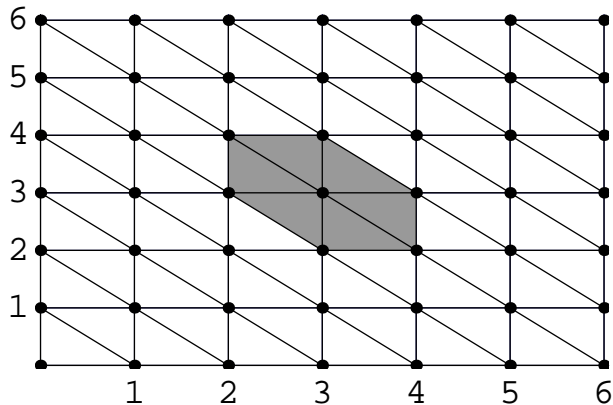}
\vspace*{-0.5cm}
\end{center}
\caption{Support of ``hexagonal'' pyramidal function for the square lattice $\Lambda=\Z^2$.}
\label{quadrilat}
\end{figure}

   A simple example is provided by the square lattice $\Lambda=\Z^2$: we just have to  divide each square in the same manner into
   two triangles. Then we get one scaling function which is a ``hexagonal pyramid'', as is shown in  Fig.~\ref{quadrilat}.

\begin{figure}[t]
\begin{center}
\includegraphics[width=13cm,clip=]{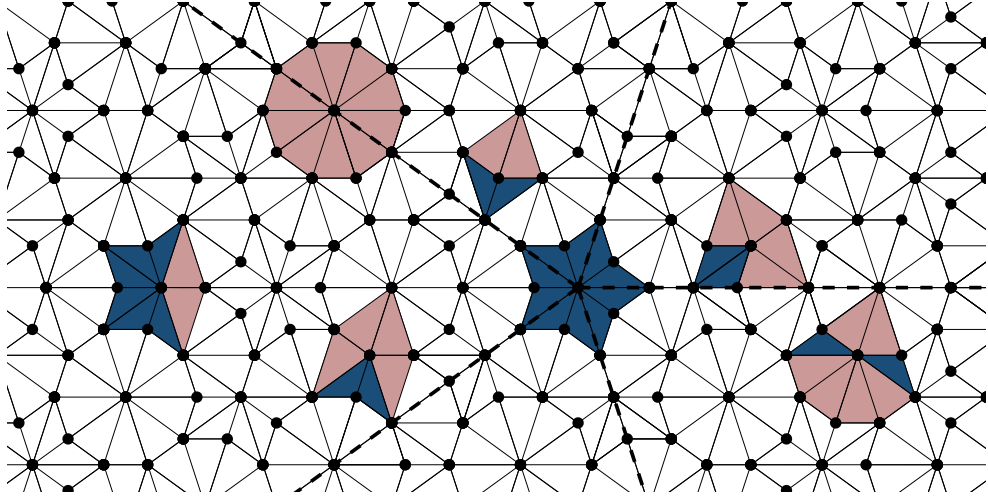}
\end{center}
\caption{Supports of seven different pyramidal functions for a (five-fold symmetrical) Penrose tiling.}
\label{penspline}
\end{figure}

Another example is provided by a five-fold Penrose tiling. Seven pyramidal functions exist here. As is shown in Fig.~\ref{penspline}, their respective supports are (in terms of kites, darts and rhombuses): a decagonal patch of  5 kites, a star-shaped patch of 5 darts, a kite-shaped patch of 2 kites and 1 ``fat'' rhombus (\emph{i.e} 2 obtuse triangles), a rhombus-shaped patch of 1 dart and 1 kite (an ``ace''), a patch of 3 darts and 1 ``thin'' rhombus (\emph{i.e} 2 acute triangles), a patch of 1 dart and 2 thin rhombuses, and a patch of 3 kites and 2 obtuse triangles.

\section{Conclusion}
The Bernuau construction of wavelets adapted to self-similar aperiodic point sets has been carried out in dealing with  some one-dimensional  examples. Our aim is currently to apply these wavelets to the analysis of aperiodic structures, like diffraction
spectra of Fibonacci chain, and to compare our results with more standard wavelet analysis ({\it e.g.} dyadic wavelets). Of course,
an essential step in  decomposition and recomposition schemes will be the determination of corresponding biorthogonal basis.   We
also plan to extend our constructions to higher-dimensional cases, like Penrose or Danzer  tilings, in view of practical
applications to quasicrystalline studies.

\subsubsection*{Acknowledgements}
{\it\small We are grateful to Dr E. Pelantov\'{a} for  helpful
  discussions. This research was partially supported by grant GACR 201/01/0130.}


\begin{thebibliography}{99}

\bibitem{mallat} S. Mallat, ``Multiresolution approximation and orthonormal bases of
wavelets for $L^2(R)$'', Trans. Amer. Math. Soc. {\bf 315}  (1989) 69--87.

\bibitem{auscher} P. Auscher,``Wavelet bases for $L^2(R)$ with rational dilation factor'',
in {\it Wavelets and Their Applications}, edited by M.B. Ruskai
{\it et al}, (Jones and Barlett), p.~439--452, 1992.

\bibitem{bumi} C. Buhmann and M. Micchelli, ``Spline prewavelets for
  nonuniform knots'', Numerische Mathematik,  {\bf 61}  (1992) 455--474.

\bibitem{dala1} X. Dai and D.R. Larson, ``Wandering vectors for unitary systems and orthogonal wavelets'', Mem. Amer. Math. Soc. {\bf 134}, n$^{\circ}$~640 (1998).

\bibitem{dala2}  X. Dai, D.R. Larson, and D. Speegle, ``Wavelets sets
  in $\R^n$'', J. Fourier Anal. Appl. {\bf 3}  (1997) 451--456.

\bibitem{waya} Y. Wang,  ``Wavelets, tilings, and spectral sets'',
  Duke Math. J. {\bf 114}, n$^{\circ}$~1,  (2002) 43--57.

\bibitem{gazpat} J.P. Gazeau and J. Patera, ``Tau wavelets of Haar'',
  J. Phys. A: Math. Gen. {\bf 29} (1996) 4549--4559.

\bibitem{bern1} G. Bernuau, {\it Propri\'et\'es spectrales et g\'eom\'etriques des quasicristaux.
Ondelettes adapt\'ees aux quasicristaux}, PhD thesis, Ceremade, Universit\'e Paris IX Dauphine,
France, 1998.

\bibitem{bern2} G. Bernuau, ``Wavelet bases adapted to a self-similar quasicrystal'', J.~Math.
Phys.  {\bf 39}  (1998) 4213-4225.

\bibitem{lem} P.G. Lemari\'e-Rieusset, ``Base d'ondelettes sur les
  groupes de Lie stratifi\'es'', Bull. Soc. Math. Fr., {\bf 117}
  (1989) 211--232.

\bibitem{abug} M. Andrle,  \v{C}. Burd\'{\i}k, J.P. Gazeau, and R. Krejcar, ``Wavelet
multiresolutions for the Fibonacci chain,'' J. Phys A: Math. Gen. {\bf
  33}  (2000) L47--L51.

\bibitem{schum} L.L. Schumaker, {\it Spline Functions: Basic Theory,} Wiley, New-York, 1981.

\bibitem{ris} J.J. Risler, {\it Math\'{e}matiques pour la CAO},
Masson, Paris, 1991.

\bibitem{gaz1} J.P. Gazeau,  ``Pisot-cyclotomic Integers for
Quasicrystals'', {\it The Mathematics of Aperiodic Long Range Order},
Ed. R.V. Moody, Nato ASI Series {\bf 382}, Kluwer, Dordrecht,  p. 175, 1997.

\bibitem{gaz2}  Ch. Frougny, J.P. Gazeau, and R. Krejcar, ``Additive and multiplicative
properties of point sets based on Beta-Integers'',Theoretical Computer
Science  {\bf 303} (2003)  491--516.

\bibitem{Me95} Y.\,Meyer, ``Quasicrystals, Diophantine approximation and
algebraic numbers'', in {\em Beyond Quasicrystals}, (F.\,Axel and D.\,
Gratias, eds), Les \'editions de physique, Springer-Verlag, 1995.

\bibitem{Mo2} R.\,Moody, ``Model Sets: A Survey'', in
{\em From Quasicrystals to More Complex Systems}, (F.\,Axel, F.\,Denoyer
and J.-P.\,Gazeau eds.), EDP Sciences and Springer Verlag, 2000.

\bibitem{Mopa} R.V. Moody and J. Patera, ``Quasicrystals and
  icosians'',   J. Phys A: Math. Gen. {\bf 26}  (1993) 2829--2853.

\bibitem{BFGK} \v C.\,Burd\'{\i}k, Ch.\,Frougny,
J.-P.\,Gazeau, and R.\,Krejcar, ``Beta-integers as natural counting
systems for quasicrystals'', {J. of Physics A: Math. Gen.} {\bf
31} (1998) 6449--6472.

\bibitem{BFGK2} \v C.\,Burd\'{\i}k, Ch.\,Frougny,
J.-P.\,Gazeau, and R.\,Krejcar, ``Beta-integers as a group'',
in {\em Dynamical Systems: From Crystal to Chaos}, World Scientific
(2000)  125--136.

\bibitem{mapape} Z.\,Mas\'akov\'a, J.\,Patera, and E. Pelantov\'a, ``Substitution rules for
aperiodic sequences of the cut and project type'', {J. of Physics A: Math. Gen.} {\bf
33} (2000) 8867--8886.

\bibitem{lest} D. Levine and P.J. Steinhardt, ``Quasicrystals I: Definitions and structure'',
{Phys. Rev. B} {\bf 34 } (1986) 596--616.

\bibitem{gazkrej} J.P. Gazeau and R. Krejcar, ``Penrose Tiling Wavelets and Quasicrystals'',
Proceedings of the Colloquium ``Complex Geometry 98'',  Eds. F.~Norguet and S. Ofman, Hermann, Paris, 2003.

\bibitem{gazkram} J.P. Gazeau and P. Kramer, ``From Quasiperiodic Tilings with $\tau$-inflation to $\tau$-wavelets'',
in {\it Proceedings of the VII$^\text{th}$ Int. Conference on Quasicrystals},
Materials Science and Engineering 294--296 (2000) 425--428.

\bibitem{ankra} M. Andrle and P. Kramer, 
{``Haar wavelets for the icosahedral Danzer tiling'',}
submitted to J. Phys. A: Math. Gen.
\end{thebibliography}
\end{document}